\documentclass[1p,times,preprint]{elsarticle}
\usepackage[english]{babel}
\usepackage{tensor}
\usepackage{graphicx}
\usepackage{amsmath}
\usepackage{amssymb}
\usepackage{amsfonts}
\usepackage{dcolumn}
\usepackage{bm}
\usepackage{xcolor}
\usepackage{ulem}
\usepackage{tikz}
\usepackage{subcaption}
\usepackage{comment}
\usepackage{verbatim}
\usepackage{fancyvrb}
\usepackage{cancel}
\usepackage{multirow}
\usepackage{longtable}
\usepackage{lscape}
\usepackage{txfonts}
\usepackage{mathtools}
\usepackage{soul}
\usepackage{url}
\usepackage{makecell}
\usepackage[pdftex]{pict2e}
\usepackage{microtype}

\def\bra<#1|{\mathinner{\langle\,{#1}\,\vert}} 
\def\ket|#1>{\mathinner{\vert\,{#1}\,\rangle}} 
\def\red|#1|{\mathinner{\!\vert\,{#1}\,\vert\!}}
\def\braket<#1>{\mathinner{\langle\,{#1}\,\rangle}} 

\def\redmem#1#2#3{  \left\langle #1 \left\Vert  
                  #2 \right\Vert #3 \right\rangle   }

\begin{document}

\title{SECOND-ORDER RAYLEIGH-SCHR\"ODINGER PERTURBATION THEORY FOR THE GRASP2018 PACKAGE: CORE-CORE CORRELATIONS}
\date{\today}

\author[TFAI]{G. Gaigalas}
\address[TFAI]{Institute of Theoretical Physics and Astronomy, 
               Vilnius University, Saul\.{e}tekio Ave. 3, LT-10257 Vilnius, Lithuania}
\ead{gediminas.gaigalas@tfai.vu.lt}

\author[TFAI]{P. Rynkun}
\ead{pavel.rynkun@tfai.vu.lt}

\author[TFAI]{L. Kitovien\.{e}}
\ead{laima.radziute@tfai.vu.lt}
 
%
%
\begin{abstract}
{\sc Grasp} package is based on the relativistic configuration interaction in which accurate calculations, accounting for valence,
valence-valence, core-valence, core, and core-core electron correlations, often rely on massive CSF expansions.
This paper presents further development of the method based on the second-order perturbation theory for finding the most important CSFs that have the greatest influence on the core-valence, core, and core-core correlations. This method is based on a combination of the relativistic configuration interaction method 
 and the stationary second-order Rayleigh-Schr\"odinger many-body perturbation theory in an irreducible tensorial form
[G. Gaigalas, P. Rynkun, L. Kitovienė, 
Second-Order Rayleigh-Schr\"odinger Perturbation Theory for the {\sc Grasp}2018 Package: Core-Valence
Correlations, {\em Lithuanian Journal of Physics}, {\bf 64}, No. 1, 20-39 (2024) (https://doi.org/10.3952/physics.2024.64.1.3)
and 
G. Gaigalas, P. Rynkun, L. Kitovienė, 
Second-Order Rayleigh-Schr\"odinger Perturbation Theory for the {\sc Grasp}2018 Package: Core
Correlations, {\em Lithuanian Journal of Physics}, {\bf 64}, No. 2, 73-81 (2024) (https://doi.org/10.3952/physics.2024.64.2.1)].
In this extension, the perturbation theory accounts for electron
core-valence, core, and core-core correlations where an atom or ion has any number of valence electrons for calculation of energy spectra and other properties.
Meanwhile the rest of the correlations are accounted for in a traditional way.
This allows a significant reduction of the space of the configuration state function
for complex atoms and ions. We also demonstrate how this method works
for calculations of the energy structure and E1 transition properties of Fe XV ion.

\end{abstract}

\maketitle

\newpage

\section{Introduction}

This paper describes the method, an extension to approach~\cite{Gaigetal:2024CV,Gaigetal:2024C}, that allows atomic structure calculations to be performed faster and with the use of less resources in the framework of relativistic atomic theory. The method is based on a combination of the relativistic configuration interaction (RCI) method~\cite{grantBook:07,Fisetal:16a} and on the stationary second-order Rayleigh-Schr\"odinger many-body perturbation theory
(for non-relativistic version of this perturbation theory in determinants, see~\cite{LindgrenBook:82})
in an irreducible tensorial form (for non-relativistic version of this perturbation theory in an irreducible tensorial form, see~\cite{Meratal:85,Meratal:86,Gaigalas:89}). In this extension, the perturbation theory accounts for electron
core-valence (CV), core (C), and core-core (CC) correlations where an atom or ion has any number of valence electrons
meanwhile the rest of the correlations are accounted for in a traditional way.

{\sc Grasp}~\cite{grasp2013,grasp2018} is based on the relativistic configuration interaction, and the wave functions of the targeted states are given as expansions over configuration state functions (CSF) built on relativistic one-electron orbitals~\cite{Peretal:23a}. Accurate RCI calculations, accounting for valence,
valence-valence, core-valence, core, and core-core electron correlations, often rely on massive CSF expansions obtained from single- and
double (SD) excitations from a multireference (MR) consisting of the most important configurations~\cite{Fisetal:16a}. 
Our proposed approach uses second-order perturbation theory to find the most important CSFs that have the greatest influence on the core-valence, core, and core-core correlations. 
This reduces the CSF base on the one hand, while on the other hand it allows the calculation of the energy structure, transition characteristics and other properties of atoms and ions to be carried out without loss of accuracy.
Since {\sc Grasp} uses the formalism of tensorial algebra~\cite{Gaigalas:2022}, only Rayleigh-Schr\"odinger  many-body perturbation theory (RSMBPT) in an irreducible tensorial form can be used in it. This makes the work unique and makes it necessary to have the expressions for the Feynman diagrams in irreducible tensorial form. It is possible to obtain them by using the combination of the angular momentum theory, the concept of irreducible tensorial sets, the generalized graphical approach, and the second quantization in coupling tensorial form~\cite{Gaigalas_1996,Gaigalas_1997}. All these expressions for the Feynman diagrams corresponding to core-core correlations in relativistic atomic theory are first presented in Section~\ref{sec:seconOrder}.
In order to be able to use the spin-angular program library~\cite{Gaigalas:2022} without any further modification, these formulas 
have first been adapted in a way that is aligned with {\sc Grasp}~\cite{grasp2013,grasp2018}. They are given in Section~\ref{sec:PT_imple}.
Meanwhile, a similar study on core-valence and core correlations 
was first done in our previous work~\cite{Gaigetal:2024CV,Gaigetal:2024C}. Both studies make full use of Racah algebra including quasispin~\cite{Gaigalas_1997} for 
integration of spin-angular part of all these types (core-valence, core, and core-core) of correlations.
The validity and efficiency of the presented method is demonstrated in Section~\ref{sec:Calculation}, where the energy spectrum and E1 transition properties of the Fe XV ion are theoretically studied.

\section{Relativistic second order effective Hamiltonian of an atom or an ion in irreducible tensorial form for core-core correlations}
\label{sec:seconOrder}

All possible core-core correlations can be treated perturbatively including all of them via RSMBPT method. All types of CC correlations are described below in separate subsections in more detail.

\subsection{The first type of core-core correlations}
\label{subsec:first_type}

Here we will discuss the first type of core-core correlations which are presented through vacuum Feynman diagrams CC$_1$ from Fig.~\ref{CC_1} and CC$_2$ from Fig.~\ref{CC_2}:
\begin{eqnarray}
\label{eq:CC1-a}
&
\hspace{-7.0cm}
(n_{a} \ell_{a})\, j_{a}^{2j_a+1} \,  (n_{b} \ell_{b})\, j_{b}^{2j_b+1} \, 
(n_{m} \ell_{m})\, j_{m}^{w_m} \, (n_{n} \ell_{n})\, j_{n}^{w_n} 
   \nonumber  \\[1ex]
&
\hspace{1.5cm}	
   \rightarrow (n_{a} \ell_{a})\, j_{a}^{2j_a} \,  (n_{b} \ell_{b})\, j_{b}^{2j_b} \,
	(n_{m} \ell_{m})\, j_{m}^{w_m} \; (n_{n} \ell_{n})\, j_{n}^{w_n} \;
	(n_{r} \ell_{r})\, j_{r} \; (n_{s} \ell_{s})\, j_{s} ,
\end{eqnarray}
\begin{eqnarray}
\label{eq:CC1-b}
&
\hspace{-7.0cm}
(n_{a} \ell_{a})\, j_{a}^{2j_a+1} \,  (n_{b} \ell_{b})\, j_{b}^{2j_b+1} \, 
(n_{m} \ell_{m})\, j_{m}^{w_m} \, (n_{n} \ell_{n})\, j_{n}^{w_n}
   \nonumber  \\[1ex]
&
\hspace{0.3cm}
   \rightarrow (n_{a} \ell_{a})\, j_{a}^{2j_a} \,  (n_{b} \ell_{b})\, j_{b}^{2j_b} \,
	(n_{m} \ell_{m})\, j_{m}^{w_m} \; (n_{n} \ell_{n})\, j_{n}^{w_n} \;
	(n_{r} \ell_{r})\, j_{r}^2 ,
\end{eqnarray}
\begin{eqnarray}
\label{eq:CC1-c}
&
\hspace{-7.2cm}
(n_{a} \ell_{a})\, j_{a}^{2j_a+1} 
(n_{m} \ell_{m})\, j_{m}^{w_m} \, (n_{n} \ell_{n})\, j_{n}^{w_n} 
   \nonumber  \\[1ex]
&
\hspace{1.6cm}
   \rightarrow (n_{a} \ell_{a})\, j_{a}^{2j_a-1} \,
	(n_{m} \ell_{m})\, j_{m}^{w_m} \; (n_{n} \ell_{n})\, j_{n}^{w_n} \;
	(n_{r} \ell_{r})\, j_{r} \;(n_{s} \ell_{s})\, j_{s} ,
\end{eqnarray}
\begin{eqnarray}
\label{eq:CC1-d}
&
\hspace{-7.2cm}
(n_{a} \ell_{a})\, j_{a}^{2j_a+1} 
(n_{m} \ell_{m})\, j_{m}^{w_m} \, (n_{n} \ell_{n})\, j_{n}^{w_n} 
   \nonumber  \\[1ex]
&
\hspace{1.6cm}
   \rightarrow (n_{a} \ell_{a})\, j_{a}^{2j_a-1} \,
	(n_{m} \ell_{m})\, j_{m}^{w_m} \; (n_{n} \ell_{n})\, j_{n}^{w_n} \;
	(n_{r} \ell_{r})\, j_{r}^2 .
\end{eqnarray}

\begin{figure*}
\begin{center}
\setlength{\unitlength}{1mm}
\begin{picture}(180,23)
\thicklines
\multiput(10,18)(1,0){10}{\circle*{0.35}}
\qbezier(10,8)(05,13)(10,18)
\put(7.6,13){\vector(0,1){2}}
\put(5.5,13){\makebox(0,0)[t]{\small{$r$}}}
\qbezier(10,8)(15,13)(10,18)
\put(12.4,13){\vector(0,-1){2}}
\put(13.5,16){\makebox(0,0)[t]{\small{$a$}}}
\qbezier(20,8)(15,13)(20,18)
\put(17.6,13){\vector(0,1){2}}
\put(16.5,10.5){\makebox(0,0)[t]{\small{$s$}}}
\qbezier(20,8)(25,13)(20,18)
\put(22.4,13){\vector(0,-1){2}}
\put(24.5,13){\makebox(0,0)[t]{\small{$b$}}}
\multiput(10,8)(1,0){10}{\circle*{0.35}}
\put(15,4){\makebox(0,0){$\text{CC}_{1}$}}
\put(27,13){\makebox(0,0) [l] {$\displaystyle{ = \frac{1}{2}
	\sum_{r, s, a, b}~\frac{\left( -1 \right)^{j_a+j_b+j_r+j_s}}{\left( \varepsilon_a+\varepsilon_b-\varepsilon_r-\varepsilon_s \right)}
\sum_{k,k'}~\frac{\delta \left( k, k' \right)}{\sqrt{\left[k,k'\right]}}
	~X_{k}(a b, r s) ~ X_{k'}(r s, a b)}$}} 

\end{picture}
\caption{The CC Feynman diagram of the second-order effective Hamiltonian for direct part of excitation 
$(n_{a} \ell_{a})\, j_{a}^{2j_a+1} \,  (n_{b} \ell_{b})\, j_{b}^{2j_b+1} \, 
(n_{m} \ell_{m})\, j_{m}^{w_m} \, (n_{n} \ell_{n})\, j_{n}^{w_n} 
   \rightarrow (n_{a} \ell_{a})\, j_{a}^{2j_a} \,  (n_{b} \ell_{b})\, j_{b}^{2j_b} \,
	(n_{m} \ell_{m})\, j_{m}^{w_m} \; (n_{n} \ell_{n})\, j_{n}^{w_n} \;
	(n_{s} \ell_{s})\, j_{s} \; (n_{r} \ell_{r})\, j_{r}$, where $a \neq b$ or $a = b$ and $s \neq r$ or $s = r$.}
\label{CC_1}
\end{center}
\end{figure*}

\begin{figure*}
\begin{center}
\setlength{\unitlength}{1mm}
\begin{picture}(180,23)
\thicklines
\multiput(10,18)(1,0){10}{\circle*{0.35}}
\put(20,18){\line(-1,-1){10}}
\put(18,10){\vector(-1,1){2}}
\put(11.5,15){\makebox(0,0)[t]{\small{$r$}}}
\put(10,8){\line(0,10){10}}
\put(10,13){\vector(0,-1){2}}
\put(7.5,13){\makebox(0,0)[t]{\small{$a$}}}
\put(20,8){\line(0,10){10}}
\put(20,13){\vector(0,-1){2}}
\put(22.5,13){\makebox(0,0)[t]{\small{$b$}}}
\put(20,8){\line(-1,1){10}}
\put(16,14){\vector(1,1){2}}
\put(11.5,12){\makebox(0,0)[t]{\small{$s$}}}
\multiput(10,8)(1,0){10}{\circle*{0.35}}
\put(15,04){\makebox(0,0){$\text{CC}_{2}$}}
\put(25,13){\makebox(0,0) [l] {$\displaystyle{ = \frac{1}{2}
  \sum_{r, s, a, b}~\frac{\left( -1 \right)^{j_a+j_{b}+j_r+j_s}}{\left( \varepsilon_{a}+\varepsilon_b-\varepsilon_r-\varepsilon_s \right)}
  \sum_{k,k'}
  \left\{
    \begin{array}{ccc}
      k  & j_{a} & j_{r} \\
      k' & j_{b} & j_{s}
    \end{array} \right\} 
	~X_{k}(a b, r s) ~ X_{k'}(s r, a b)}$}}
\end{picture}
\caption{The CC Feynman diagram of the second-order effective Hamiltonian for exchange part of excitation
$    (n_{a} \ell_{a})\, j_{a}^{2j_a+1} \,  (n_{b} \ell_{b})\, j_{b}^{2j_b+1} \, 
(n_{m} \ell_{m})\, j_{m}^{w_m} \, (n_{n} \ell_{n})\, j_{n}^{w_n} 
   \rightarrow (n_{a} \ell_{a})\, j_{a}^{2j_a} \,  (n_{b} \ell_{b})\, j_{b}^{2j_b} \,
	(n_{m} \ell_{m})\, j_{m}^{w_m} \; (n_{n} \ell_{n})\, j_{n}^{w_n} \;
	(n_{s} \ell_{s})\, j_{s} \; (n_{r} \ell_{r})\, j_{r}$, where $a \neq b$ or $a = b$ and $s \neq r$ or $s = r$.}
\label{CC_2}
\end{center}
\end{figure*}

These two Feynman diagrams CC$_1$ and CC$_2$ are vacuum diagrams without any spin-angular part therefore spin-angular program library~\cite{Gaigalas:2022} from {\sc Grasp}
is not needed. They have the same impact on those CSFs that have the same configuration. 
The diagram CC$_1$ describes a direct part of excitations (\ref{eq:CC1-a})-(\ref{eq:CC1-d}) and CC$_2$ describes an exchange part of the same excitations.

Each second order Feynman diagram's expression of Rayleigh-Schr\"odinger  many-body perturbation theory as described in~\cite{Gaigetal:2024CV}
 has the energy denominator
$D = \sum \left( \varepsilon_{\text{down}} - \varepsilon_{\text{up}} \right)$,
where $\varepsilon_{\text{down}}$ ($\varepsilon_{\text{up}}$) is the single-particle eigenvalue associated with the down- (up-)
orbital lines to (from) the lowest interaction line of the diagram. For example the denominators for  CC$_1$ and CC$_2$ diagrams are
\begin{equation}
\label{eq:denominator}
D = \left( \varepsilon_{a}+\varepsilon_b-\varepsilon_r-\varepsilon_s \right),
\end{equation}
where indexes $a$ and $b$ belong to $F$ set, and $r$, $s$ belong to $G$ set of orbitals~\cite{Gaigetal:2024CV}. 

Also, the following notations are used in the expressions of these diagrams (see Fig.~\ref{CC_1} and Fig.~\ref{CC_2}):
\begin{equation}
\label{eq:deffX}
   X_{k}(i j, i' j') 
   = \redmem{\ell_i j_{i}}{\, C^{(k)} \,}{ \ell_{i'} j_{i'}}
     \redmem{\ell_j j_{j}}{\, C^{(k)} \,}{ \ell_{j'} j_{j'}} 
R^{k}(n_i j_i \, n_jj_j, \, n_{i'}j_{i'} \, n_{j'}j_{j'} ) ,
\end{equation}
where $R^{k}\left(n_i j_i \, n_jj_j, \, n_{i'}j_{i'} \, n_{j'}j_{j'} \right)$ is the radial integral 
of electrostatic interaction between electrons~\cite[(89) and (90)]{Fisetal:16a} and 
$\redmem{\ell_i j_{i}}{\, C^{(k)} \,}{ \ell_{i'} j_{i'}}$ is the reduced matrix element of the irreducible tensor operator $C^{(k)}$ in $jj$-coupling.

The exchange diagram CC$_2$ additionally has $6j$- coefficient. The summation in the expresions of CC$_1$ and CC$_2$ are running over closed lines of the Feynman diagrams and over the ranks $k$ and $k'$ of the irreducible tensor operators $C^{(k)}$ and $C^{(k')}$, respectively.

\subsection{The second type of core-core correlations}
\label{subsec:second_type}

\begin{figure*}
\begin{center}
\setlength{\unitlength}{1mm}
\begin{picture}(180,33)
\thicklines
\multiput(10,30)(1,0){10}{\circle*{0.35}}
\qbezier(10,20)(05,25)(10,30)
\put(7.6,25){\vector(0,1){2}}
\put(5.5,25){\makebox(0,0)[t]{\small{$r$}}}
\qbezier(10,20)(15,25)(10,30)
\put(12.4,25){\vector(0,-1){2}}
\put(14,25){\makebox(0,0)[t]{\small{$a$}}}
\put(20,30){\line(1,-1){10}}
\put(28,28){\vector(1,1){2.4}}
\put(27,27){\vector(1,1){2}}
\put(27,31){\makebox(0,0)[t]{\small{$m$}}}
\put(20,20){\line(0,10){10}}
\put(20,25){\vector(0,-1){2}}
\put(18,25.6){\makebox(0,0)[t]{\small{$b$}}}
\put(20,20){\line(1,1){10}}
\put(29,21){\vector(-1,1){2.4}}
\put(26.2,20.4){\makebox(0,0){\small{$m'$}}}
\put(30,20){\vector(-1,1){2}}
\multiput(10,20)(1,0){10}{\circle*{0.35}}
\put(15,16){\makebox(0,0){$\text{CC}_{3}$}}
\put(33,25){\makebox(0,0) [l] {$\displaystyle{ =
\sum_{m, m^{\prime}}~\frac{1}{\sqrt{\left[j_m \right]}}
\left[\;  \tilde a^{\left( j_{m^{\prime}} \right) }  \times
  a^{\left( j_{m} \right) } \; \right] ^{\left( 0 \right)} 
	\sum_{r, a, b}~\frac{\left( -1 \right)^{j_a+j_b+j_r+j_m}}{\left( \varepsilon_a+\varepsilon_b-\varepsilon_r-\varepsilon_m \right)}
}$}}
\put(43,10){\makebox(0,0) [l] {$\displaystyle{ \times
\sum_{k,k'}~~\frac{\delta \left( k, k' \right)}{\sqrt{\left[k,k'\right]}}
	~X_{k}(a b, r m^{\prime}) ~ X_{k'}(r m, a b)}$}} 

\end{picture}
\caption{The CC Feynman diagram of the second-order effective Hamiltonian for direct part of excitation 
$(n_{a} \ell_{a})\, j_{a}^{2j_a+1} \,  (n_{b} \ell_{b})\, j_{b}^{2j_b+1} \, 
(n_{m} \ell_{m})\, j_{m}^{w_m} \, (n_{n} \ell_{n})\, j_{n}^{w_n} 
   \rightarrow (n_{a} \ell_{a})\, j_{a}^{2j_a} \,  (n_{b} \ell_{b})\, j_{b}^{2j_b} \,
	(n_{m} \ell_{m})\, j_{m}^{w_m+1} \; (n_{n} \ell_{n})\, j_{n}^{w_n} \;
	(n_{r} \ell_{r})\, j_{r}$, where $a \neq b$ or $a = b$.}
\label{CC_3}
\end{center}
\end{figure*}

\begin{figure*}
\begin{center}
\setlength{\unitlength}{1mm}
\begin{picture}(180,33)
\thicklines
\multiput(10,30)(1,0){10}{\circle*{0.35}}
\put(20,30){\line(-1,-1){10}}
\put(17,23){\vector(1,-1){2}}
\put(19.5,27){\makebox(0,0)[t]{\small{$a$}}}
\put(10,20){\line(0,10){10}}
\put(10,25){\vector(0,1){2}}
\put(7.5,25){\makebox(0,0)[t]{\small{$r$}}}
\put(20,30){\line(1,-1){10}}
\put(28,28){\vector(1,1){2.4}}
\put(27,27){\vector(1,1){2}}
\put(27,31){\makebox(0,0)[t]{\small{$m$}}}
\put(20,20){\line(-1,1){10}}
\put(18,28){\vector(-1,-1){2}}
\put(19.5,24.5){\makebox(0,0)[t]{\small{$b$}}}
\put(20,20){\line(1,1){10}}
\put(29,21){\vector(-1,1){2.4}}
\put(26.2,20.4){\makebox(0,0){\small{$m'$}}}
\put(30,20){\vector(-1,1){2}}
\multiput(10,20)(1,0){10}{\circle*{0.35}}
\put(15,16){\makebox(0,0){$\text{CC}_{4}$}}
\put(33,25){\makebox(0,0) [l] {$\displaystyle{ =
\sum_{m, m^{\prime}}~\frac{1}{\sqrt{\left[j_m \right]}}
\left[\;  \tilde a^{\left( j_{m^{\prime}} \right) }  \times
  a^{\left( j_{m} \right) } \; \right] ^{\left( 0 \right)} 
  \sum_{r, a, b}~\frac{\left( -1 \right)^{j_a+j_{b}+j_r+j_m}}{\left( \varepsilon_{a}+\varepsilon_b-\varepsilon_r-\varepsilon_m \right)}
	}$}}
\put(43,10){\makebox(0,0) [l] {$\displaystyle{ \times
  \sum_{k,k'}
  \left\{
    \begin{array}{ccc}
      k  & j_{a} & j_{m} \\
      k' & j_{b} & j_{r}
    \end{array} \right\} 
	~X_{k}(b a, r m^{\prime}) ~ X_{k'}(r m, a b)}$}}
\end{picture}
\caption{The CC Feynman diagram of the second-order effective Hamiltonian for exchange part of excitation $(n_{a} \ell_{a})\, j_{a}^{2j_a+1} \,  (n_{b} \ell_{b})\, j_{b}^{2j_b+1} \, 
(n_{m} \ell_{m})\, j_{m}^{w_m} \, (n_{n} \ell_{n})\, j_{n}^{w_n} 
   \rightarrow (n_{a} \ell_{a})\, j_{a}^{2j_a} \,  (n_{b} \ell_{b})\, j_{b}^{2j_b} \,
	(n_{m} \ell_{m})\, j_{m}^{w_m+1} \; (n_{n} \ell_{n})\, j_{n}^{w_n} \;
	(n_{r} \ell_{r})\, j_{r}$, where $a \neq b$ or $a = b$.}
\label{CC_4}
\end{center}
\end{figure*}

This type of core-core correlations is presented through the Feynman diagrams CC$_3$ from Fig.~\ref{CC_3} and CC$_4$ from Fig.~\ref{CC_4} where all lines with double arrow of diagrams are renamed $m$, i.e. $m' \equiv m$:

\begin{eqnarray}
\label{eq:CC2-a}
&
\hspace{-7.0cm}
(n_{a} \ell_{a})\, j_{a}^{2j_a+1} \,  (n_{b} \ell_{b})\, j_{b}^{2j_b+1} \, 
(n_{m} \ell_{m})\, j_{m}^{w_m} \, (n_{n} \ell_{n})\, j_{n}^{w_n} 
   \nonumber  \\[1ex]
&
\hspace{1.5cm}	
   \rightarrow (n_{a} \ell_{a})\, j_{a}^{2j_a} \,  (n_{b} \ell_{b})\, j_{b}^{2j_b} \,
	(n_{m} \ell_{m})\, j_{m}^{w_m+1} \; (n_{n} \ell_{n})\, j_{n}^{w_n} \;
	(n_{r} \ell_{r})\, j_{r} ,
\end{eqnarray}
\begin{eqnarray}
\label{eq:CC2-b}
&
\hspace{-7.2cm}
(n_{a} \ell_{a})\, j_{a}^{2j_a+1} \,
(n_{m} \ell_{m})\, j_{m}^{w_m} \, (n_{n} \ell_{n})\, j_{n}^{w_n} 
   \nonumber  \\[1ex]
&
\hspace{1.6cm}
   \rightarrow (n_{a} \ell_{a})\, j_{a}^{2j_a-1} \,
	(n_{m} \ell_{m})\, j_{m}^{w_m+1} \; (n_{n} \ell_{n})\, j_{n}^{w_n} \;
	(n_{r} \ell_{r})\, j_{r} .
\end{eqnarray}

These diagrams CC$_3$ and CC$_4$ are single particle Feynman diagrams where their tensorial part is expressed through tensorial product of annihilation $\tilde a^{\left( j\right) }$ and creation $a^{\left( j \right) }$ operators
\begin{equation}
\label{eq:s-a_CC3}
 \left[\; \tilde a^{\left( j_{m^{\prime}}  \right) } \times
   a^{\left( j_m \right) } \; \right] ^{\left( 0 \right)} .
\end{equation}
These two operators of second quantization act on the same subshell $m$ and represent the scalar operator. Therefore, this operator can be expressed through the operator of subshell occupation number $\hat{N}$ on which it acts
\begin{equation}
\label{eq:s-b_CC3}
 \left[\; \tilde a^{\left( j_{m}  \right) } \times
   a^{\left( j_m \right) } \; \right] ^{\left( 0 \right)} = \frac{\left[ j_m \right] - \hat{N}_m}{\sqrt{\left[ j_m \right]}}
\end{equation}
or through the hole operator  $\hat{N}_{\text{hol}}$ of this subshell
\begin{equation}
\label{eq:s-c_CC3}
 \left[\; \tilde a^{\left( j_{m}  \right) } \times
   a^{\left( j_m \right) } \; \right] ^{\left( 0 \right)} = \frac{ \hat{N}_{\text{hol}~m}}{\sqrt{\left[ j_m \right]}} .
\end{equation}
Therefore, as for the first type of core-core correlation (see Subsection~\ref{subsec:first_type}), there is no need to use the spin-angular program library~\cite{Gaigalas:2022} to calculate these diagrams, as the spin-angular coefficients are expressed through a simple multiplier (see diagram A3 in~\cite{Gaigetal:2005}). This is one of the advantages of the methodology proposed in this paper.

\subsection{The third type of core-core correlations}
\label{subsec:third_type}

\begin{figure*}
\begin{center}
\setlength{\unitlength}{1mm}
\begin{picture}(180,33)
\thicklines
\multiput(20,30)(1,0){10}{\circle*{0.35}}
\put(30,30){\line(-1,-1){10}}
\put(26,24){\vector(1,-1){2}}
\put(21.5,27){\makebox(0,0)[t]{\small{$a$}}}
\put(30,20){\line(0,10){10}}
\put(30,23.5){\vector(0,1){2}}
\put(30,25.5){\vector(0,1){2}}
\put(32.5,25){\makebox(0,0)[t]{\small{$n$}}}
\put(10,30){\line(1,-1){10}}
\put(11,21){\vector(1,1){2.4}}
\put(10,20){\vector(1,1){2}}
\put(13.2,31){\makebox(0,0)[t]{\small{$m$}}}
\put(30,20){\line(-1,1){10}}
\put(28,28){\vector(-1,-1){2}}
\put(21.5,24.9){\makebox(0,0)[t]{\small{$b$}}}
\put(10,20){\line(1,1){10}}
\put(12,28){\vector(-1,1){2.4}}
\put(13.5,20.4){\makebox(0,0){\small{$m'$}}}
\put(13,27){\vector(-1,1){2}}
\multiput(20,20)(1,0){10}{\circle*{0.35}}
\put(25,16){\makebox(0,0){$\text{CC}_{5}$}}
\put(36,25){\makebox(0,0) [l] {$\displaystyle{ =
 \frac{1}{2} \sum_{m, m^{\prime}}~\frac{1}{\sqrt{\left[j_m \right]}}
\left[\;  \tilde a^{\left( j_{m^{\prime}} \right) }  \times
  a^{\left( j_{m} \right) } \; \right] ^{\left( 0 \right)} 
  \sum_{n, a, b}~\frac{\left( -1 \right)^{j_a+j_{b}+j_m+j_n}}{\left( \varepsilon_{a}+\varepsilon_b-\varepsilon_m-\varepsilon_n \right)}
	}$}}
\put(46,10){\makebox(0,0) [l] {$\displaystyle{ \times
  \sum_{k,k'}
  \left\{
    \begin{array}{ccc}
      k  & j_{a} & j_{m} \\
      k' & j_{b} & j_{n}
    \end{array} \right\} 
	~X_{k}(a b, m^{\prime} n) ~ X_{k'}(m n, b a)}$}}
\end{picture}
\caption{The CC one-particle Feynman diagram of the second-order effective Hamiltonian for 
the third type 
$(n_{a} \ell_{a})\, j_{a}^{2j_a+1} \, (n_{b} \ell_{b})\, j_{b}^{2j_b+1} \, (n_{m} \ell_{m})\, j_{m}^{w_m} \, (n_{n} \ell_{n})\, j_{n}^{w_n} 
   \rightarrow (n_{a} \ell_{a})\, j_{a}^{2j_a} \; (n_{b} \ell_{b})\, j_{b}^{2j_b} \;(n_{m} \ell_{m})\, j_{m}^{w_m+2} \; (n_{n} \ell_{n})\, j_{n}^{w_n}$  and the fourth type
$(n_{a} \ell_{a})\, j_{a}^{2j_a+1} \,  (n_{b} \ell_{b})\, j_{b}^{2j_b+1} \, 
(n_{m} \ell_{m})\, j_{m}^{w_m} \, (n_{n} \ell_{n})\, j_{n}^{w_n}
   \rightarrow (n_{a} \ell_{a})\, j_{a}^{2j_a} \,  (n_{b} \ell_{b})\, j_{b}^{2j_b} \,
	(n_{m} \ell_{m})\, j_{m}^{w_m+1} \; (n_{n} \ell_{n})\, j_{n}^{w_n+1}$
of core-core correlations.}
\label{CC_5}
\end{center}
\end{figure*}

\begin{figure*}
\begin{center}
\setlength{\unitlength}{1mm}
\begin{picture}(180,48)
\thicklines
\multiput(20,45)(1,0){10}{\circle*{0.35}}
\put(10,45){\line(1,-1){10}}
\put(11,36){\vector(1,1){2.4}}
\put(10,35){\vector(1,1){2}}
\put(13.2,46){\makebox(0,0)[t]{\small{$m$}}}
\put(10,35){\line(1,1){10}}
\put(12,43){\vector(-1,1){2.4}}
\put(13.5,35.4){\makebox(0,0){\small{$m'$}}}
\put(13,42){\vector(-1,1){2}}
\put(20,35){\line(0,10){10}}
\put(20,40){\vector(0,-1){2}}
\put(22.5,40){\makebox(0,0)[t]{\small{$a$}}}
\put(30,35){\line(0,10){10}}
\put(30,40){\vector(0,-1){2}}
\put(27.5,40.7){\makebox(0,0)[t]{\small{$b$}}}
\put(30,45){\line(1,-1){10}}
\put(38,43){\vector(1,1){2.4}}
\put(37,42){\vector(1,1){2}}
\put(37,46){\makebox(0,0)[t]{\small{$n$}}}

\put(30,35){\line(1,1){10}}
\put(39,36){\vector(-1,1){2.4}}
\put(36.2,35.4){\makebox(0,0){\small{$n'$}}}
\put(40,35){\vector(-1,1){2}}
\multiput(20,35)(1,0){10}{\circle*{0.35}}
\put(25,31){\makebox(0,0){$\text{CC}_{6}$}}
\put(43,40){\makebox(0,0) [l] {$\displaystyle{ = 
	\frac{1}{2} \sum_{m, m^{\prime}}~\sum_{n, n^{\prime}}~\sum_{j_{12}}~~\sqrt{\left[ j_{12} \right]}~
	\left[\left[\; \tilde a^{\left( j_{m^{\prime}} \right) }  \times 
  a^{\left( j_m \right) } \; \right] ^{\left( j_{12} \right)} \times 
	\left[\; \tilde a^{\left( j_{n^{\prime}} \right) }  \times 
  a^{\left( j_n \right) } \; \right] ^{\left( j_{12} \right)} \right]^{\left( 0 \right)}
	}$}}
\put(53,25){\makebox(0,0) [l] {$\displaystyle{ \times
  \sum_{a, b}~\frac{1}{\left( \varepsilon_{a}+\varepsilon_b-\varepsilon_m-\varepsilon_n \right)}
  \sum_{k,k'}
  \left\{
    \begin{array}{ccc}
      k & k' & j_{12} \\
      m & m' & j_{a}
    \end{array} \right\}
  \left\{
    \begin{array}{ccc}
      k & k' & j_{12} \\
      n & n' & j_{b}
    \end{array} \right\}
		}$}}
\put(53,10){\makebox(0,0) [l] {$\times \; X_{k}(a b, m^{\prime} n^{\prime}) ~ X_{k'}(m n, a b)$}}
\end{picture}
\caption{The CC two-particle Feynman diagram of the second-order effective Hamiltonian for 
the third type 
$(n_{a} \ell_{a})\, j_{a}^{2j_a+1} \, (n_{b} \ell_{b})\, j_{b}^{2j_b+1} \, (n_{m} \ell_{m})\, j_{m}^{w_m} \, (n_{n} \ell_{n})\, j_{n}^{w_n} 
   \rightarrow (n_{a} \ell_{a})\, j_{a}^{2j_a} \; (n_{b} \ell_{b})\, j_{b}^{2j_b} \;(n_{m} \ell_{m})\, j_{m}^{w_m+2} \; (n_{n} \ell_{n})\, j_{n}^{w_n}$  and the fourth type
$(n_{a} \ell_{a})\, j_{a}^{2j_a+1} \,  (n_{b} \ell_{b})\, j_{b}^{2j_b+1} \, 
(n_{m} \ell_{m})\, j_{m}^{w_m} \, (n_{n} \ell_{n})\, j_{n}^{w_n}
   \rightarrow (n_{a} \ell_{a})\, j_{a}^{2j_a} \,  (n_{b} \ell_{b})\, j_{b}^{2j_b} \,
	(n_{m} \ell_{m})\, j_{m}^{w_m+1} \; (n_{n} \ell_{n})\, j_{n}^{w_n+1}$
of core-core correlations.}
\label{CC_6}
\end{center}
\end{figure*}

This type of core-core correlations is presented through one-particle Feynman diagram CC$_5$  from Fig.~\ref{CC_5} and through two-particle Feynman diagram CC$_6$ from Fig.~\ref{CC_6} where all lines with double arrow of diagrams are renamed $m$, i.e. $m' \equiv m$ for CC$_5$ and $m' = n = n' \equiv m$  for CC$_6$:

\begin{eqnarray}
\label{eq:CC3-a}
&
\hspace{-7.0cm}
(n_{a} \ell_{a})\, j_{a}^{2j_a+1} \,  (n_{b} \ell_{b})\, j_{b}^{2j_b+1} \, 
(n_{m} \ell_{m})\, j_{m}^{w_m} \, (n_{n} \ell_{n})\, j_{n}^{w_n} 
   \nonumber  \\[1ex]
&
\hspace{1.5cm}	
   \rightarrow (n_{a} \ell_{a})\, j_{a}^{2j_a} \,  (n_{b} \ell_{b})\, j_{b}^{2j_b} \,
	(n_{m} \ell_{m})\, j_{m}^{w_m+2} \; (n_{n} \ell_{n})\, j_{n}^{w_n} ,
\end{eqnarray}
\begin{eqnarray}
\label{eq:CC3-b}
&
\hspace{-7.2cm}
(n_{a} \ell_{a})\, j_{a}^{2j_a+1} \,
(n_{m} \ell_{m})\, j_{m}^{w_m} \, (n_{n} \ell_{n})\, j_{n}^{w_n} 
   \nonumber  \\[1ex]
&
\hspace{1.6cm}
   \rightarrow (n_{a} \ell_{a})\, j_{a}^{2j_a-1} \,
	(n_{m} \ell_{m})\, j_{m}^{w_m+2} \; (n_{n} \ell_{n})\, j_{n}^{w_n} .
\end{eqnarray}

The CC$_5$ diagram has the same tensorial part as the diagrams CC$_3$ and CC$_4$ therefore its spin-angular part can be handled in the same way as for CC$_3$ and CC$_4$ diagrams.

The CC$_6$ is a two-particle Feynman diagram where its tensorial part is expressed through tensorial product of annihilation $\tilde a^{\left( j\right) }$ and creation $a^{\left( j \right) }$ operators and represent the scalar operator for this type of core-core correlations
\begin{eqnarray}
\label{eq:CC3-c}
	\left[\left[\; \tilde a^{\left( j_{m^{\prime}} \right) }  \times 
  a^{\left( j_m \right) } \; \right] ^{\left( j_{12} \right)} \times 
	\left[\; \tilde a^{\left( j_{n^{\prime}} \right) }  \times 
  a^{\left( j_n \right) } \; \right] ^{\left( j_{12} \right)} \right]^{\left( 0 \right)} .
\end{eqnarray}
For this CC$_6$ diagram, unlike for the other Feynman diagrams considered above, there comes the additive summation, where the summation parameter $j_{12}$ relates the tensor product (\ref{eq:CC3-c}) to the two $6j$- coefficients. This summation parameter is the intermediate rank of this tensor product.

All operators of second quantization act on the same subshell $m$ therefore the tensorial operator
(\ref{eq:CC3-c}) can be expressed through simple $\hat{N}$ and $\hat{N}_{\text{hol}}$ operators in case $j_{12}=0$
\begin{eqnarray}
\label{eq:CC3-d}
	\left[\left[\; \tilde a^{\left( j_m \right) }  \times 
  a^{\left( j_m \right) } \; \right] ^{\left( 0 \right)} \times 
	\left[\; \tilde a^{\left( j_m \right) }  \times 
  a^{\left( j_m \right) } \; \right] ^{\left( 0 \right)} \right]^{\left( 0 \right)}
	= \frac{\left( \left[ j_m \right] - \hat{N}_m \right)^2}{\left[ j_m \right]}
\end{eqnarray}
or
\begin{eqnarray}
\label{eq:CC3-e}
	\left[\left[\; \tilde a^{\left( j_{m} \right) }  \times 
  a^{\left( j_m \right) } \; \right] ^{\left( 0 \right)} \times 
	\left[\; \tilde a^{\left( j_{m} \right) }  \times 
  a^{\left( j_m \right) } \; \right] ^{\left( 0 \right)} \right]^{\left( 0 \right)}
	 = \frac{ \hat{N}^2_{\text{hol}~m}}{\left[ j_m \right]} .
\end{eqnarray}
So for this case, for calculation of CC$_6$ diagram, we also do not
need to use the spin-angular program library~\cite{Gaigalas:2022}. Meanwhile,
the program library~\cite{Gaigalas:2022} from {\sc Grasp} supports calculation of spin-angular part (\ref{eq:CC3-c}) of CC$_6$ Feynman diagram in case $j_{12}>0$ after some simple modifications similar to~\cite{Gaigalas_1996} will be performed in Section~\ref{sec:PT_imple}. After this modification, full Racah algebra including quasispin~\cite{Gaigalas_1997} is available for integration of spin-angular part of this Feynman diagram.
This is another advantage of the methodology proposed in this paper.

\subsection{The fourth type  of core-core correlations}
\label{subsec:fourth_type}

This type of core-core correlations is presented through the Feynman diagram CC$_6$ where all lines with double arrow of diagrams are renamed in the following way: $m' \equiv m$ and $n' \equiv n$

\begin{eqnarray}
\label{eq:CC4-a}
&
\hspace{-7.0cm}
(n_{a} \ell_{a})\, j_{a}^{2j_a+1} \,  (n_{b} \ell_{b})\, j_{b}^{2j_b+1} \, 
(n_{m} \ell_{m})\, j_{m}^{w_m} \, (n_{n} \ell_{n})\, j_{n}^{w_n} 
   \nonumber  \\[1ex]
&
\hspace{1.5cm}	
   \rightarrow (n_{a} \ell_{a})\, j_{a}^{2j_a} \,  (n_{b} \ell_{b})\, j_{b}^{2j_b} \,
	(n_{m} \ell_{m})\, j_{m}^{w_m+1} \; (n_{n} \ell_{n})\, j_{n}^{w_n+1} ,
\end{eqnarray}
\begin{eqnarray}
\label{eq:CC4-b}
&
\hspace{-7.0cm}
(n_{a} \ell_{a})\, j_{a}^{2j_a+1} \,  
(n_{m} \ell_{m})\, j_{m}^{w_m} \, (n_{n} \ell_{n})\, j_{n}^{w_n} 
   \nonumber  \\[1ex]
&
\hspace{1.5cm}	
   \rightarrow (n_{a} \ell_{a})\, j_{a}^{2j_a-1} \,  
	(n_{m} \ell_{m})\, j_{m}^{w_m+1} \; (n_{n} \ell_{n})\, j_{n}^{w_n+1} .
\end{eqnarray}
In this case the tensorial part of the CC$_6$ diagram has the following form:
\begin{eqnarray}
\label{eq:CC4-c}
	\left[\left[\; \tilde a^{\left( j_m \right) }  \times 
  a^{\left( j_m \right) } \; \right] ^{\left( j_{12} \right)} \times 
	\left[\; \tilde a^{\left( j_n \right) }  \times 
  a^{\left( j_n \right) } \; \right] ^{\left( j_{12} \right)} \right]^{\left( 0 \right)} ,
\end{eqnarray}
which can be expressed in case $j_{12} = 0$  as
\begin{eqnarray}
\label{eq:CC4-d}
	\left[\left[\; \tilde a^{\left( j_m \right) }  \times 
  a^{\left( j_m \right) } \; \right] ^{\left( 0 \right)} \times 
	\left[\; \tilde a^{\left( j_n \right) }  \times 
  a^{\left( j_n \right) } \; \right] ^{\left( 0 \right)} \right]^{\left( 0 \right)}
	= \frac{\left( \left[ j_m \right] - \hat{N}_m \right) \; \left( \left[ j_n \right] - \hat{N}_n \right)}{\sqrt{\left[ j_m, j_n \right]}}
\end{eqnarray}
or
\begin{eqnarray}
\label{eq:CC4-e}
	\left[\left[\; \tilde a^{\left( j_m \right) }  \times 
  a^{\left( j_m \right) } \; \right] ^{\left( 0 \right)} \times 
	\left[\; \tilde a^{\left( j_n \right) }  \times 
  a^{\left( j_n \right) } \; \right] ^{\left( 0 \right)} \right]^{\left( 0 \right)}
	= \frac{ \hat{N}_{\text{hol}~m} \; \hat{N}_{\text{hol}~n}}{\sqrt{\left[ j_m, j_n \right]}} .
\end{eqnarray}
So for this case we also do not
need to use the spin-angular program library~\cite{Gaigalas:2022} for calculation of the CC$_6$ diagram. Meanwhile,
the program library~\cite{Gaigalas:2022} from {\sc Grasp} supports calculation of the spin-angular part (\ref{eq:CC4-c}) of the CC$_6$ Feynman diagram in case $j_{12}>0$ after some simple modifications similar to ~\cite{Gaigalas_1996} will be performed in Section~\ref{sec:PT_imple}. After this modification, full Racah algebra including quasispin~\cite{Gaigalas_1997} is available for integration of the spin-angular part of this Feynman diagram.

\subsection{The contribution of core-core correlations to off-diagonal matrix elements}
\label{subsec:offdiagonal_type}

The main contribution of core-core correlations to off-diagonal matrix elements is in the matrix element
$\redmem{(n_{m} \ell_{m})\, j_{m}^{w_m} \, (n_{n} \ell_{n})\, j_{n}^{w_n}}{\, \widehat{{\cal H}}^{(2)}_{\text{Effective}} \,}{(n_{m} \ell_{m})\, j_{m}^{w_m-2} \, (n_{n} \ell_{n})\, j_{n}^{w_n+2} }$.
The above contribution is derived from the excitation:
\begin{equation}
\label{eq:CC-off_Diagonal_a} 
(n_{a} \ell_{a}) \; j_{a}^{2j_a+1} \, (n_{b} \ell_{b}) \; j_{b}^{2j_b+1} \; (n_{m} \ell_{m})\; j_{m}^{w_m} \; (n_{n} \ell_{n})\; j_{n}^{w_n}
   \rightarrow (n_{a} \ell_{a})\, j_{a}^{2j_a} \; (n_{b} \ell_{b}) \; j_{b}^{2j_b} \; (n_{m} \ell_{m}) \; j_{m}^{w_m} \; (n_{n} \ell_{n}) \; j_{n}^{w_n+2}
\end{equation}
and can be described through the same two-particle Feynman diagram Fig.~\ref{CC_6} as above.
This type of core-core correlations is presented through the Feynman diagram CC$_6$ where all lines with double arrow of diagrams are renamed in the following way: $m \equiv n$, $m' \equiv m$, and $n' \equiv m$.

In this case, the tensorial part of the CC$_6$ diagram has the following form:
\begin{eqnarray}
\label{eq:CC-off_Diagonal_b}
	\left[\left[\; \tilde a^{\left( j_m \right) }  \times 
  a^{\left( j_n \right) } \; \right] ^{\left( j_{12} \right)} \times 
	\left[\; \tilde a^{\left( j_m \right) }  \times 
  a^{\left( j_n \right) } \; \right] ^{\left( j_{12} \right)} \right]^{\left( 0 \right)} .
\end{eqnarray}
In this case, after some modifications made in Section~\ref{sec:PT_imple}, the Racah algebra~\cite{Gaigalas_1997} and the software
library~\cite{Gaigalas:2022} are also fully available.

\section{PT implementation in the {\sc Grasp}-2018~\cite{grasp2013,grasp2018}}
\label{sec:PT_imple}

Similar to CV and C correlations \cite{Gaigetal:2024CV,Gaigetal:2024C}, the admixed configurations from CC correlations can be added to usual energy $E_0 \left(K \right)$ of the 
term $\chi J$ of the configuration $K$ and can be
expressed as the energy $\Delta \mathcal{E}_0 \left(K J \right)$, which does not depend on the term, and the sum of the product of Slater integrals and spin-angular coefficients, describing the interaction within open subshells and between them:
\begin{eqnarray}
\label{eq:BogEnergy}
\hspace*{-2.5cm}
   E\left(K \chi J \right)
	\nonumber \\
& &
   = E_0 \left(K J\right) + \Delta \mathcal{E}_0 \left(K J \right) 
	\nonumber \\  [0.2cm]
& &
	+ \; \sum_{n\ell j} \sum_{k>0} \widetilde{f}_k \left( \ell j^{w}, \; K \chi J  \right)
	\left[ \mathcal{F}^{k} \left( n \ell j, \; n \ell j \right)  
	+ \Delta \mathcal{F}^{k} \left( n \ell j, \; n \ell j \right) \right]
	\nonumber \\
& &
	+ \; \sum_{n\ell j} \sum_{n'\ell'j' > n\ell j} \left\{ \sum_{k>0} \widetilde{f}_k \left( \ell j^{w} \; \ell' j'^{w'},
	\; K \chi J  \right) \right.
\left[ \mathcal{F}^{k} \left( n \ell j, \; n '\ell' j' \right)  
	+ \Delta \mathcal{F}^{k} \left( n \ell j, \; n' \ell' j' \right) \right]
	\nonumber \\ [0.2cm]
& & 
	+ \sum_{k} \widetilde{g}_k \left( \ell j^{w} \; \ell' j'^{w'}, \; K \chi J  \right)	
\left[ \mathcal{G}^{k} \left( n \ell j, \; n '\ell' j' \right)  
	+ \Delta \mathcal{G}^{k} \left( n \ell j, \; n' \ell' j' \right) \right] 
	\nonumber \\ [0.2cm]
& &
	+ \sum_{k} \widetilde{v}_k \left( \ell j^{w} \; \ell' j'^{w'}, \ell j^{w-2} \; \ell' j'^{w'+2},
	\; K \chi J \; K' \chi' J \right)	
	\nonumber \\
& &
  \times 
\left. 
\left[ \mathcal{R}^{k} \left( n \ell j n \ell j, \; n '\ell' j' n '\ell' j' \right)  
	+ \Delta \mathcal{R}^{k} \left( n \ell j n \ell j, \; n' \ell' j' n '\ell' j' \right) \right] \right\} ,
\end{eqnarray}
where $\widetilde{f}_k$, $\widetilde{g}_k$, and $\widetilde{v}_k$ are spin-angular coefficients from which submatrix elements $\redmem{\ell j}{\, C^{(k)} \,}{ \ell^{\prime} j^{\prime}}$ are extracted. 
Therefore summation over $k$ runs over all 
possible values instead of the values which satisfy the triangular condition $\left( \ell \ell^{\prime} k\right)$ as it is in the ordinary case. The $\mathcal{F}^{k} \left( n \ell j, \; n '\ell' j' \right)$,
$\mathcal{G}^{k} \left( n \ell j, \; n '\ell' j' \right)$, and $\mathcal{R}^{k} \left( n \ell j n \ell j, \; n '\ell' j' n '\ell' j' \right)$ are generalized integrals of electrostatic interaction between electrons. The definition of $\mathcal{R}^{k} \left( n \ell j n \ell j, \; n '\ell' j' n '\ell' j' \right)$ is the following:
\begin{eqnarray}
\label{eq:BogRk}
\hspace*{-2.5cm}
   \mathcal{R}^{k}\left(i j, i' j'\right)
	\nonumber \\
& &
   = \left\{ \left[ 1 + \delta \left( i, j \right) \right]  \left[ 1 + \delta \left( i', j' \right) \right]  \right\} ^{-1/2}
	 \, R^{k}\left(n_i j_i \, n_jj_j, \, n_{i'}j_{i'} \, n_{j'}j_{j'} \right)
\nonumber \\
& &
	   \times \redmem{\ell_i j_{i}}{\, C^{(k)} \,}{ \ell_{i'} j_{i'}}
     \redmem{\ell_j j_{j}}{\, C^{(k)} \,}{ \ell_{j'} j_{j'}}, 
\end{eqnarray}
where $R^{k}\left(n_i j_i \, n_jj_j, \, n_{i'}j_{i'} \, n_{j'}j_{j'} \right)$ is the same radial integral as in 
Eq. (\ref{eq:deffX}). Definitions  $\mathcal{F}^{k} \left( n \ell j, \; n '\ell' j' \right)$,
$\mathcal{G}^{k} \left( n \ell j, \; n '\ell' j' \right)$ straightforwardly follow from Eq. (\ref{eq:BogRk}).

The contribution coming from the CC correlations of the configurations $K'$ to $E (K \chi J)$ in the second order of the perturbation theory
can be written from Eq. (\ref{eq:BogEnergy}) as
\begin{eqnarray}
\label{eq:BogEnergy_PT}
\hspace*{-2.5cm}
  \Delta E_{PT (CC)}
	\nonumber \\
& &
   =  \Delta \mathcal{E}_0 \left(K J \right) 
	\nonumber \\  [0.2cm]
& &
	+ \; \sum_{n\ell j} \sum_{k>0} \widetilde{f}_k \left( \ell j^{w}, \; K \chi J  \right)
	\Delta \mathcal{F}^{k} \left( n \ell j, \; n \ell j \right)
	\nonumber \\
& &
	+ \; \sum_{n\ell j} \sum_{n'\ell'j' > n\ell j} \left\{ \sum_{k>0} \widetilde{f}_k \left( \ell j^{w} \; \ell' j'^{w'},
	\; K \chi J  \right) \right.
 \Delta \mathcal{F}^{k} \left( n \ell j, \; n' \ell' j' \right) 
	\nonumber \\ [0.2cm]
& & 
	+ \sum_{k} \widetilde{g}_k \left( \ell j^{w} \; \ell' j'^{w'}, \; K \chi J  \right)	
 \Delta \mathcal{G}^{k} \left( n \ell j, \; n' \ell' j' \right)
	\nonumber \\ [0.2cm]
& &
\left.
	+ \sum_{k} \widetilde{v}_k \left( \ell j^{w} \; \ell' j'^{w'}, \ell j^{w-2} \; \ell' j'^{w'+2},
	\; K \chi J \; K' \chi' J \right)	
\Delta \mathcal{R}^{k} \left( n \ell j n \ell j, \; n' \ell' j' n '\ell' j' \right) \right\} .
\end{eqnarray}

\begin{table*}
\begin{center}
\begin{tabular}{|l|} \hline
$\Delta \mathcal{E}_0$ corrections \\ \hline \hline
\\
$\overbrace{(n_{a} \ell_{a})\, j_{a}^{2j_a+1} \; (n_{b} \ell_{b})\, j_{b}^{2j_b+1}}^{\text{core subshells}} \, \overbrace{(n_{m} \ell_{m})\, j_{m}^{w_m} \; (n_{n} \ell_{n})\, j_{n}^{w_n}}^{\text{valence subshells}} \;
\rightarrow \; \overbrace{(n_{a} \ell_{a})\, j_{a}^{2j_a} \; (n_{b} \ell_{b})\, j_{b}^{2j_b}}^{\text{core subshells}} \; \overbrace{(n_{m} \ell_{m})\, j_{m}^{w_m} \; (n_{n} \ell_{n})\, j_{n}^{w_n}}^{\text{valence subshells}} \; \overbrace{(n_{r} \ell_{r})\, j_{r} \; (n_{s} \ell_{s})\, j_{s}}^{\text{virtual subshells}}$ \\
\\
$	- \underbrace{\sqrt{\left[ j_a\right]} \left[ \sqrt{\left[ j_r\right]} \; \mathcal{A}\left( 0, \, a b, \, s r \right)
	+ \sqrt{\left[ j_s\right]} \; \mathcal{A}\left( 0, \, a b, \, r s \right) \right]}_{\text{from $\text{CC}_1$ Feynman diagram}} 
-\underbrace{(-1)^{j_a+j_b+j_r+j_s}\sum_{k} \, \left[
  \mathcal{C}\left(k, \, a b, \, r s \right) + \mathcal{C}\left(k, \, b a , \, r s \right) \right]}_{\text{from $\text{CC}_2$ Feynman diagram}}$
\\
\\ \hline \hline
\\
$ \overbrace{(n_{a} \ell_{a})\, j_{a}^{2j_a+1} \; (n_{b} \ell_{b})\, j_{b}^{2j_b+1}}^{\text{core subshells}} \, \overbrace{(n_{m} \ell_{m})\, j_{m}^{w_m} \; (n_{n} \ell_{n})\, j_{n}^{w_n}}^{\text{valence subshells}} \;
\rightarrow  \; \overbrace{(n_{a} \ell_{a})\, j_{a}^{2j_a} \; (n_{b} \ell_{b})\, j_{b}^{2j_b}}^{\text{core subshells}} \; \overbrace{(n_{m} \ell_{m})\, j_{m}^{w_m+1} \; (n_{n} \ell_{n})\, j_{n}^{w_n}}^{\text{valence subshells}} \; 
\overbrace{(n_{r} \ell_{r})\, j_{r}}^{\text{virtual subshells}}$ \\
\\
$\frac{w_m - \left[j_m\right]} {\left[j_m \right]}
\left\{ \underbrace{\sqrt{\left[ j_m\right]} \left[ \sqrt{\left[ j_a\right]} \; \mathcal{A}\left( 0, \, a b, \, r m \right)
	+ \sqrt{\left[ j_b\right]} \; \mathcal{A}\left( 0, \, b a, \, r m \right) \right]}_{\text{from $\text{CC}_3$ Feynman diagram}} \right.$ \\ 
 \\ \hspace{1.9cm} $\left.
+ \underbrace{(-1)^{j_a+j_b+j_r+j_m}\sum_{k} \, \left[
  \mathcal{C}\left(k, \, a b, \, r m \right) + \mathcal{C}\left(k, \, b a , \, r m \right) \right]}_{\text{from $\text{CC}_4$ Feynman diagram}} \right\}$ \\
\\ \hline 
\end{tabular}
\end{center}
\caption{Expressions for the first (where $a \neq b$ or $a = b$ and $s \neq r$ or $s = r$) and second (where $a \neq b$ or $a = b$) types of core-core corrections to the energy in Eq. (\ref{eq:BogEnergy}), independent of the term.}
\label{tab:Implemen_CC1}
\end{table*}

The contribution of the first two types of CC correlations in the second-order of the perturbation theory is expressed only over $\Delta \mathcal{E}_0 \left(K J \right)$ (see Table~\ref{tab:Implemen_CC1}). 
The contributions $\Delta \mathcal{F}^{k} \left( n \ell j, \; n \ell j \right)$, $\Delta \mathcal{F}^{k} \left( n \ell j, \; n' \ell' j' \right)$,
and $\Delta \mathcal{G}^{k} \left( n \ell j, \; n' \ell' j' \right)$ are equal to zero in these cases. 
The contributions $\Delta \mathcal{E}_0 \left(K J \right)$ can be expressed through $\mathcal{A}$ and $\mathcal{C}$ coefficients (see Table~\ref{tab:Implemen_CC1}) which have the following expressions:

\begin{equation}
\label{eq:BogA}
   \mathcal{A}\left(x, \; i j, \; i' j'\right) 
  = \sum_{k,k'}
		  \left\{
    \begin{array}{ccc}
      k  & k' & x \\
      j_{i} & j_{i} & j_{i'}
    \end{array} \right\}
				  \left\{
    \begin{array}{ccc}
      k  & k' & x \\
      j_{j'} & j_{j'} & j_{j}
    \end{array} \right\}
\mathcal{P}\left(kk', \; i j, \; i' j'\right) ,
\end{equation}

\begin{equation}
\label{eq:BogC}
   \mathcal{C}\left(x, \; i j, \; i' j'\right) 
  = \sum_{k}
		  \left\{
    \begin{array}{ccc}
      x  & j_{i} & j_{i'} \\
      k & j_{j} & j_{j'}
    \end{array} \right\}
\mathcal{Q}\left(xk, \; i j, \; i' j'\right) ,
\end{equation}

where
\begin{equation}
\label{eq:BogP}
   \mathcal{P}\left(kk', \; i j, \; i' j'\right) 
   = \mathcal{R}^{k}\left(i j, \; i' j'\right) \; \mathcal{R}^{k'}\left(i' j', \; i j \right)  \; 
	\mathcal{O}\left(K', K \right),
\end{equation}
\begin{equation}
\label{eq:BogQ}
   \mathcal{Q}\left(kk', \; i j, \; i' j'\right)
   = \mathcal{R}^{k}\left(i j, \; i' j'\right) \; \mathcal{R}^{k'}\left(i' j', \; j i\right)  \; 
\mathcal{O}\left(K', K \right) ,
\end{equation}
and
\begin{equation}
\label{eq:BogO1}
\mathcal{O}\left(K', K \right)
= \frac{1}{\overline{E}\left(K' \right)-\overline{E}\left(K\right)},
\end{equation}
where $\overline{E}\left(K\right)$ is the averaged energy of the 
state 
for which calculations are performed. $\overline{E}\left(K^{'}\right)$ is the averaged energy for the admixed configuration $K'$.
For details on how to find $\overline{E}\left(K\right)$ and $\overline{E}\left(K^{'}\right)$, see Section~3 in~\cite{Gaigetal:2024CV}.
We would like to emphasize that the energy denominator (\ref{eq:BogO1}) is defined differently/opposite to the expressions of Feynman diagrams 
(see for example Fig.~\ref{CC_1}, Eqs. (\ref{eq:BogP}), (\ref{eq:BogQ})).

As we can see, the expression of the second type of core-core correlations (see Table~\ref{tab:Implemen_CC1}), unlike the first type of core-core correlations, depends on the occupation number $w_m$ of subshell $m$ (number of electrons in the subshell $m$). 
This is due to the fact that the second type of core-core correlations is described by the CC$_3$ and CC$_4$ Feynman diagrams, whose angular part is expressed via the operator $\hat{N}$ of subshell occupation number (see Eq. (\ref{eq:s-b_CC3})). Meanwhile, the first type of core-core correlations is described by vacuum Feynman diagrams CC$_1$ and CC$_2$ which do not have any tensorial part in their expressions (see Subsection~\ref{subsec:first_type}).

\begin{table*}
\begin{center}
\begin{tabular}{|l|} \hline
$\Delta \mathcal{E}_0$ corrections \\ \hline \hline
\\
$\overbrace{(n_{a} \ell_{a})\, j_{a}^{2j_a+1} \; (n_{b} \ell_{b})\, j_{b}^{2j_b+1}}^{\text{core subshells}} \, \overbrace{(n_{m} \ell_{m})\, j_{m}^{w_m} \; (n_{n} \ell_{n})\, j_{n}^{w_n}}^{\text{valence subshells}} \;
\rightarrow \; \overbrace{(n_{a} \ell_{a})\, j_{a}^{2j_a} \; (n_{b} \ell_{b})\, j_{b}^{2j_b}}^{\text{core subshells}} \; \overbrace{(n_{m} \ell_{m})\, j_{m}^{w_m+2}\; (n_{n} \ell_{n})\, j_{n}^{w_n}}^{\text{valence subshells}}$ \\
\\ 
$\underbrace{\; -2 \; \frac{\left(\left[ j_m \right] - w_m \right)\left(\left[ j_m \right] - w_m - 1 \right)} {\left[ j_m \right]}
 \,
\mathcal{A}^{\prime}\left( 0, \, m m, \, a b \right)
+ 
\; 2 \; \frac{\left[ j_m \right] - 2 w_m}{\left[ j_m \right]}
\sum_{k>0} \left( -1 \right)^{k} \left[ k \right] \mathcal{A}^{\prime} \left( k, \, m m, \, a b \right) }_{\text{from $\text{CC}_{5}$ and $\text{CC}_{6}$ Feynman diagrams}}$ \\
\\ \hline \hline
 \\
$\overbrace{(n_{a} \ell_{a})\, j_{a}^{2j_a+1} \; (n_{b} \ell_{b})\, j_{b}^{2j_b+1}}^{\text{core subshells}} \, \overbrace{(n_{m} \ell_{m})\, j_{m}^{w_m} \; (n_{n} \ell_{n})\, j_{n}^{w_n}}^{\text{valence subshells}} \;
\rightarrow \; \overbrace{(n_{a} \ell_{a})\, j_{a}^{2j_a} \; (n_{b} \ell_{b})\, j_{b}^{2j_b}}^{\text{core subshells}} \; \overbrace{(n_{m} \ell_{m})\, j_{m}^{w_m+1}\; (n_{n} \ell_{n})\, j_{n}^{w_n+1}}^{\text{valence subshells}}$ \\
\\
$\underbrace{- \; \frac{\left( \left[j_m\right]- w_m \right) \left( \left[j_n\right]-w_n\right)} {\sqrt{\left[j_m, j_n\right]}}
 \,
\left( \mathcal{A}^{\prime}\left( 0, \, m n, \, a b \right) + \mathcal{A}^{\prime}\left( 0, \, m n, \, b a \right)\right)}_{\text{from $\text{CC}_{5}$ and $\text{CC}_{6}$ Feynman diagrams}}$ 
\\ \\ \\ \hspace{1.9cm}
$\underbrace{- \frac{(-1)^{j_m+j_n+j_a+j_b}}{2} \; \left( \frac{\left( \left[j_m\right]- 2w_m \right)}{\left[j_m\right]} + \frac{\left( \left[j_n\right]-2 w_n\right)}{\left[j_n\right]} \right)
 \,
\sum_{k} \left( \mathcal{C}\left( k, \, a b, \, m n \right) + \mathcal{C}\left( k, \, a b, \, n m \right)\right) }_{\text{from $\text{CC}_{5}$ and $\text{CC}_{6}$ Feynman diagrams}}$ \\
\\ \hline 
\end{tabular}
\end{center}
\caption{Expressions for the third (where $a \neq b$ or $a = b$) and
fourth (where $a \neq b$ or $a = b$) types of core-core corrections to the energy in Eq. (\ref{eq:BogEnergy}), independent of the term.}
\label{tab:Implemen_CC2}
\end{table*}

The contribution of the last two types of CC correlations in the second-order of the perturbation theory is expressed over $\Delta \mathcal{E}_0 \left(K J \right)$, $\Delta \mathcal{F}^{k} \left( n \ell j, \; n \ell j \right)$, $\Delta \mathcal{F}^{k} \left( n \ell j, \; n' \ell' j' \right)$,
and $\Delta \mathcal{G}^{k} \left( n \ell j, \; n' \ell' j' \right)$ (see 
Tables~\ref{tab:Implemen_CC2} and \ref{tab:Implemen_CC3}).
These formulae are additionally expressed via the quantities

\begin{equation}
\label{eq:BogAp}
   \mathcal{A^{\prime}}\left(x, \; i j, \; i' j'\right) 
  = \sum_{k,k'}
		  \left\{
    \begin{array}{ccc}
      k  & k' & x \\
      j_{i} & j_{i} & j_{i'}
    \end{array} \right\}
				  \left\{
    \begin{array}{ccc}
      k  & k' & x \\
      j_{j} & j_{j} & j_{j'}
    \end{array} \right\}
\mathcal{P}\left(kk', \; i j, \; i' j'\right) ,
\end{equation}

\begin{equation}
\label{eq:BogB}
   \mathcal{B}\left(x, \; i j, \; i' j'\right) 
  = \sum_{k,k'}
		  \left\{
    \begin{array}{ccc}
      k  & k' & x \\
      j_{j} & j_{i} & j_{i'}
    \end{array} \right\}
				  \left\{
    \begin{array}{ccc}
      k  & k' & x \\
      j_{i} & j_{j} & j_{j'}
    \end{array} \right\}
\mathcal{Q}\left(kk', \; i j, \; i' j'\right) .
\end{equation}

As we can see, the expression of the last two types of core-core correlations (see Table~\ref{tab:Implemen_CC2}) for the contribution $\Delta \mathcal{E}_0$ depends on different combinations
$\frac{ \left[j\right] \; - \; w} {\sqrt{\left[j\right]}}$ of the subshells $m$ and $n$. 
This is due to the fact that these types of core-core correlations are described by the CC$_5$ and CC$_6$ Feynman diagrams, whose angular part is expressed via the combination 
$\frac{\left[ j \right] \; - \; \hat{N}}{\sqrt{\left[ j \right]}}$
of the operator acting on $n$ and/or $m$ subshell (see Subsections~\ref{subsec:third_type} and \ref{subsec:fourth_type}). 

\begin{table*}
\begin{center}
\begin{tabular}{|lll|} \hline
Corrections & Slater integral & $k$ values\\ \hline  \hline
& & \\
\multicolumn{3}{|c|}{
$\overbrace{(n_{a} \ell_{a})\, j_{a}^{2j_a+1} \; (n_{b} \ell_{b})\, j_{b}^{2j_b+1}}^{\text{core subshells}} \, \overbrace{(n_{m} \ell_{m})\, j_{m}^{w_m} \; (n_{n} \ell_{n})\, j_{n}^{w_n}}^{\text{valence subshells}} \;
\rightarrow \; \overbrace{(n_{a} \ell_{a})\, j_{a}^{2j_a} \; (n_{b} \ell_{b})\, j_{b}^{2j_b}}^{\text{core subshells}} \; \overbrace{(n_{m} \ell_{m})\, j_{m}^{w_m+2}\; (n_{n} \ell_{n})\, j_{n}^{w_n}}^{\text{valence subshells}}$} \\
& &\\
$\underbrace{ 
-4 \; \left[ k \right] \mathcal{A^{\prime}}\left( k, \, m m, \, a b \right) }_{\text{from $\text{CC}_{6}$ Feynman diagram}}$ \hspace{6cm} 
&$\Delta \mathcal{F}^{k}(m,m)$ & $k>0$\\
& &\\ \hline \hline
 & &\\
\multicolumn{3}{|c|}{
$\overbrace{(n_{a} \ell_{a})\, j_{a}^{2j_a+1} \; (n_{b} \ell_{b})\, j_{b}^{2j_b+1}}^{\text{core subshells}} \, \overbrace{(n_{m} \ell_{m})\, j_{m}^{w_m} \; (n_{n} \ell_{n})\, j_{n}^{w_n}}^{\text{valence subshells}} \;
\rightarrow \; \overbrace{(n_{a} \ell_{a})\, j_{a}^{2j_a} \; (n_{b} \ell_{b})\, j_{b}^{2j_b}}^{\text{core subshells}} \; \overbrace{(n_{m} \ell_{m})\, j_{m}^{w_m+1}\; (n_{n} \ell_{n})\, j_{n}^{w_n+1}}^{\text{valence subshells}}$} \\
& &\\
$\underbrace{ 
- \; \left[ k \right] \left( \mathcal{A^{\prime}}\left( k, \, m n, \, a b \right) + \mathcal{A^{\prime}}\left( k, \, m n, \, b a \right) \right)}_{\text{from $\text{CC}_{6}$ Feynman diagram}}$ 
&$\Delta \mathcal{F}^{k}(m,n)$ & $k>0$\\
& &\\
& &\\
$\underbrace{ 
- 2\; \left[ k \right] \mathcal{B}\left( k, \, m n, \, a b \right)}_{\text{from $\text{CC}_{6}$ Feynman diagram}}$ 
&$\Delta \mathcal{G}^{k}(m,n)$ & $k\geq 0$\\
& & \\ \hline 
\end{tabular}
\end{center}
\caption{Expressions for Slate integrals $\Delta \mathcal{F}^{k}(m,m)$, $\Delta \mathcal{F}^{k}(m,n)$ and $\Delta \mathcal{G}^{k}(m,n)$  (see Eq. (\ref{eq:BogEnergy})) corresponding to the third type 
$(n_{a} \ell_{a})\, j_{a}^{2j_a+1} \, (n_{b} \ell_{b})\, j_{b}^{2j_b+1} \, (n_{m} \ell_{m})\, j_{m}^{w_m} \, (n_{n} \ell_{n})\, j_{n}^{w_n} 
   \rightarrow (n_{a} \ell_{a})\, j_{a}^{2j_a} \; (n_{b} \ell_{b})\, j_{b}^{2j_b} \;(n_{m} \ell_{m})\, j_{m}^{w_m+2} \; (n_{n} \ell_{n})\, j_{n}^{w_n}$	(where $a \neq b$ or $a = b$) and the fourth type
$(n_{a} \ell_{a})\, j_{a}^{2j_a+1} \,  (n_{b} \ell_{b})\, j_{b}^{2j_b+1} \, 
(n_{m} \ell_{m})\, j_{m}^{w_m} \, (n_{n} \ell_{n})\, j_{n}^{w_n}
   \rightarrow (n_{a} \ell_{a})\, j_{a}^{2j_a} \,  (n_{b} \ell_{b})\, j_{b}^{2j_b} \,
	(n_{m} \ell_{m})\, j_{m}^{w_m+1} \; (n_{n} \ell_{n})\, j_{n}^{w_n+1}$ (where $a \neq b$ or $a = b$)
of core-core correlations.} 
\label{tab:Implemen_CC3}
\end{table*}

\begin{table*}
\begin{center}
\begin{tabular}{|lll|} \hline
Corrections & Slater integral & $k$ values\\ \hline  \hline
& & \\ 
$- 4 \left[ k \right] 
\underbrace{ \mathcal{X}\left( k, \; n n, \; a b, \; m m \right)}_{\text{from $\text{CC}_6$ Feynman diagram}}$ \hspace{2cm} &$\Delta \mathcal{R}^{k}(m m,nn)$ & $k \geq 0$ \\ 
& & \\ \hline
\end{tabular}
\end{center}
\caption{Expressions for Slater integral $\Delta \mathcal{R}^{k}(m m,nn)$  (see Eq. (\ref{eq:BogEnergy})) corresponding to the
core-core \\ 
$(n_{a} \ell_{a})\, j_{a}^{2j_a+1} \, (n_{b} \ell_{b})\, j_{b}^{2j_b+1} \, (n_{m} \ell_{m})\, j_{m}^{w_m} \, (n_{n} \ell_{n})\, j_{n}^{w_n} 
   \rightarrow (n_{a} \ell_{a})\, j_{a}^{2j_a} \; (n_{b} \ell_{b})\, j_{b}^{2j_b} \;(n_{m} \ell_{m})\, j_{m}^{w_m} \; (n_{n} \ell_{n})\, j_{n}^{w_n+2}$
  correlations
 coming from the off diagonal matrix element 
$\redmem{(n_{m} \ell_{m})\, j_{m}^{w_m} \, (n_{n} \ell_{n})\, j_{n}^{w_n}}{\, \widehat{{\cal H}}^{(2)}_{\text{Effective}} \,}{(n_{m} \ell_{m})\, j_{m}^{w_m-2} \, (n_{n} \ell_{n})\, j_{n}^{w_n+2} }$.} 
\label{tab:Implemen_CC_Off}
\end{table*}

The contribution of CC correlation in the second-order of the perturbation theory coming from off diagonal matrix element 
$\redmem{(n_{m} \ell_{m})\, j_{m}^{w_m} \, (n_{n} \ell_{n})\, j_{n}^{w_n}}{\, \widehat{{\cal H}}^{(2)}_{\text{Effective}} \,}{(n_{m} \ell_{m})\, j_{m}^{w_m-2} \, (n_{n} \ell_{n})\, j_{n}^{w_n+2} }$
is described by the diagram CC$_6$. Reformulation of the expressions of this diagram into the form suitable for the {\sc Grasp} gave these corrections only to the radial integral $\Delta \mathcal{R}^{k}(m m,nn)$ (see Table~\ref{tab:Implemen_CC_Off}). This formula is expressed via the quantity: 

\begin{equation}
\label{eq:BogX}
   \mathcal{X}\left(x, \; i j, \; i' j', \; i'' j''\right) 
  = \sum_{k,k'}
				  \left\{
    \begin{array}{ccc}
      k  & k' & x \\
      j_{i''} & j_{i} & j_{i'}
    \end{array} \right\}				  \left\{
    \begin{array}{ccc}
      k  & k' & x \\
      j_{j''} & j_{j} & j_{j'}
    \end{array} \right\}
\mathcal{S'}\left(kk', \; i j, \; i' j', \; i'' j''\right) ,
\end{equation}
where
\begin{equation}
\label{eq:BogSprim}
   \mathcal{S'}\left(kk', \; i j, \; i' j', \; i'' j''\right) 
  =
\mathcal{R}^{k}\left(i j, \; i' j'\right)  \; 
\mathcal{R}^{k'}\left(i' j', \; i'' j'' \right)  \; 
\mathcal{O}\left(K', K_1 K_2 \right)
\end{equation}
and
\begin{equation}
\label{eq:BogO2}
\mathcal{O}\left(K', K_1 K_2 \right)
=
\frac{1}{2} \left( {\frac{1}{\overline{E}\left(K' \right)-\overline{E}\left(K_1\right)}}
+{\frac{1}{\overline{E}\left(K' \right)-\overline{E}\left(K_2\right)}}
\right),
\end{equation}
where $\overline{E}\left(K_1\right)$ corresponds to the  averaged  energy of the configuration $K_1$ from bra function of off diagonal matrix element and  $\overline{E}\left(K_2\right)$ corresponds to the  averaged  energy of the configuration $K_1$ from ket function of off diagonal matrix element. For details on how to find them, see \cite[Section~3]{{Gaigetal:2024CV}}.

This theory in irreducible tensorial form is more suitable to be included in such version of the {\sc Grasp} which is based on configuration state function generators
(CSFGs)~\cite{grasp2023}. This is related to the fact that this version of the package allows us to distinguish $F$, $F'$, and $G$ sets of orbitals very easily in the process of computing atomic data. In the following section we will present a test case of this implementation.

\section{Calculation of Core-valence, Core, and Core-core Correlations with a New Approach}
\label{sec:Calculation}

The method, which is based on the Rayleigh-Schr\"odinger perturbation theory in an irreducible tensorial form \cite{Gaigetal:2024CV,Gaigetal:2024C},
has now been extended to include CC correlations. 
In this work, the RSMBPT method is used to include CV, C and CC correlations 
for the calculation of the Fe XV energy structure and E1 transition properties.  
The energy levels of the $\mathrm{3s^2}$, $\mathrm{3p^2}$, $\mathrm{3s3d}$, $\mathrm{3d^2}$, 
$\mathrm{3p3d}$, $\mathrm{3s3p}$ configurations and the E1 transition properties between states of these configurations are computed. 
The results are compared with those obtained by the regular RCI in the GRASP \cite{GRASP_manual}. 
The radial wave functions are taken from the earlier study \cite{Gaigetal:2024C} where 
the computational scheme is described in detail. 
The RCI calculations are performed, including the Coulomb and Breit interactions
and leading quantum electrodynamic effects – the vacuum polarization and self-energy corrections.
Meanwhile the estimation of correlations was done using the stationary second-order Rayleigh-Schr\"odinger many-body perturbation theory in an irreducible tensorial form for Coulomb interaction.

Regular RCI results are marked as \textbf{RCI} with the correlations, which were included in the calculations, 
e.g. valence-valence (VV) (\textbf{VV RCI}).
In this computational scheme 
single-double (SD) substitutions are allowed from the 3s, $\mathrm{3p_-}$, 3p, $\mathrm{3d_-}$, 3d valence orbitals 
to orbital set (OS) $OS_1$, ..., $OS_5$ (see Ref. \cite{Gaigetal:2024C}) .
In \textbf{VV+CV+C RCI} computational scheme 
SD substitutions are allowed from the 3s, $\mathrm{3p_-}$, 3p, $\mathrm{3d_-}$, 3d valence orbitals
and single (S) substitutions are allowed only from the 2s or $\mathrm{2p_-}$ and 2p core orbital 
to orbital set $OS_1$, ..., $OS_5$.
In \textbf{VV+CV+C+CC RCI} computational scheme 
SD substitutions are allowed from the 3s, $\mathrm{3p_-}$, 3p, $\mathrm{3d_-}$, 3d valence orbitals
and from the 2s, $\mathrm{2p_-}$ and 2p core orbitals to orbital set $OS_1$, ..., $OS_5$.
The multi-reference (MR) set consists of the $\mathrm{3s^2}$, $\mathrm{3p^2}$, $\mathrm{3s3d}$, $\mathrm{3d^2}$ even and 
$\mathrm{3p3d}$, $\mathrm{3s3p}$ odd configurations. 
The strategies marked as 'int' mean that only CSFs that have non-zero
matrix elements in the sets of spin-angular integration with the CSFs belonging to the configurations
in the MR are retained.

The CSF space in RSMBPT computations is divided into three sets: $F$, $F'$ and $G$ 
(see Ref. \cite{Gaigetal:2024CV} for details). Thus, the 1s is defined as inactive core subshell, 2s, $\mathrm{2p_-}$ 
and 2p subshells are defined as active core subshells 
(that correspond to $F$ set), 3s, $\mathrm{3p_-}$, 3p, $\mathrm{3d_-}$, and 3d as valence subshells 
(that correspond to $F'$ set), and subshells belonging to $OS_1$, ..., $OS_5$ as virtual ones 
(that correspond to $G$ set). 
This distribution of space is consistent with regular {\sc Grasp}2018 calculations
and allows the use of a combination of RCI and RSMBPT methods.
Results including CV, C and CC correlations according to the RSMBPT method are marked as \textbf{CV+C+CC RCI (RSMBPT)}.
In this case, the contribution of each $K'$ configuration of the CV, C and CC correlations for CSF 
for which energy needs to be calculated according to Rayleigh-Schr\"odinger perturbation theory 
in an irreducible tensorial form according 
to the Eq. (22) of Ref. \cite{Gaigetal:2024CV}, Eq. (6) of Ref. \cite{Gaigetal:2024C}, and Eq. (\ref{eq:BogEnergy_PT}) is computed.
The total contribution of the CV, C and CC correlations is also computed.
$K'$ configurations are sorted in descending order according to the impact of the CV, C and CC correlations for each level.

Further, $K'$ configurations are selected by the CV, C and CC correlations impact
with the specified fraction (expressed in the percentage: 95, 99, 99.5, 99.95 and 100\%) 
of the total CV, C and CC contribution, and RCI computations are performed including them.
These CSFs bases are also reduced by removing from the list the CSFs that have zero 
spin-angular coefficients for matrix elements with the CSFs 
belonging to the configurations in the MR. Later RCI computations are performed for reduced CSFs base.
These results are marked as 'int' in the further description.
Note that the program gives the contribution of the CV, C, and CC correlations of $K'$ configuration 
with a value greater than {\tt 1.0E-11}, smaller contributions are neglected. 
The C and CV correlations (Eqs. (3) and (4) of Ref. \cite{Gaigetal:2024C}) which are not included with RSMBPT method, were added to RCI calculations in a regular way, together with the valence and valence-valence correlations.

\subsection{Calculations of the energy structure}
\label{sec:Calculation_Energy}

Table \ref{regular_Grasp} presents the comparison for 35 energy levels from regular {\sc Grasp}2018 calculations 
with NIST ASD \cite{NIST_ASD}. In the regular {\sc Grasp}2018 calculations VV, CV, C, and CC electron correlations are included.
The number of CSFs ($N_{CSFs}$) from each calculation and the root-mean-square (rms) with NIST data
are given in the last lines of the table. 
It is seen from the table that the CV and C correlations are important, the rms deviation obtained for the energy levels
from the NIST data is 653.16 cm$^{-1}$.
By including CC correlations the rms deviation decreases to 401.16 cm$^{-1}$. 

\begin{table*}[!ht]
\setlength{\tabcolsep}{2pt}
{\scriptsize
\caption{The energy levels (in cm$^{-1}$) and differences (in cm$^{-1}$) between \textbf{RCI} and NIST
 energies ($\Delta E_{\textbf{(RCI)-(NIST)}}$) for Fe~XV are given 
when computations are performed in regular way including VV, CV, C, and CC correlations.}
\label{regular_Grasp}
\centering
\begin{tabular}{r l r r r r r r r}
\hline\hline
\multicolumn{1}{c}{\multirow{2}{*}{No.}} & \multicolumn{1}{c}{\multirow{2}{*}{State}} 
 & \multicolumn{1}{c}{NIST}&& \multicolumn{5}{c}{$\Delta E_{\textbf{(RCI)-(NIST)}}$} \\
\cline{5-9}
&&&& \multicolumn{1}{c}{VV} & \multicolumn{1}{c}{VV+CV+C int} & \multicolumn{1}{c}{VV+CV+C } & \multicolumn{1}{c}{VV+CV+C+CC int} & 
\multicolumn{1}{c}{VV+CV+C+CC} \\
\hline
\noalign{\smallskip}
 1& $\mathrm{3s^2~^1S_0}$    &      0  &&          &          &  &        &           \\      
 2& $\mathrm{3s3p~^3P^o_0}$  & 233842  &&  $-$746.91 &  8.17  &  7.82   &  $-$296.32 &  $-$301.68  \\
 3& $\mathrm{3s3p~^3P^o_1}$  & 239660  &&  $-$717.57 &  27.93 &  27.91  &  $-$285.97 &  $-$284.90  \\
 4& $\mathrm{3s3p~^3P^o_2}$  & 253820  &&  $-$798.57 &  23.34 &  23.28  &  $-$306.04 &  $-$303.93  \\
 5& $\mathrm{3s3p~^1P^o_1}$  & 351911  &&  2925.87 &  211.45  &  211.19 &  $-$8.53   &  $-$12.67   \\  
 6& $\mathrm{3p^2~^3P_0}$    & 554524  &&  1959.20 &  156.46  &  156.46 &  450.57  &  444.78   \\      
 7& $\mathrm{3p^2~^1D_2}$    & 559600  &&  286.67  &  295.66  &  295.62 &  342.25  &  341.33   \\      
 8& $\mathrm{3p^2~^3P_1}$    & 564602  &&  1840.29 &  121.10  &  121.05 &  401.73  &  394.75   \\      
 9& $\mathrm{3p^2~^3P_2}$    & 581803  &&  1442.72 &  173.83  &  173.81 &  375.78  &  373.70   \\      
10& $\mathrm{3p^2~^1S_0}$    & 659627  &&  3045.02 &  511.51  &  511.51 &  715.39  &  703.60   \\      
11& $\mathrm{3s3d~^3D_1}$    & 678772  &&  1684.31 &  331.01  &  329.60 &  68.52   &  52.91    \\      
12& $\mathrm{3s3d~^3D_2}$    & 679785  &&  1672.03 &  350.25  &  349.51 &  81.77   &  73.14    \\      
13& $\mathrm{3s3d~^3D_3}$    & 681416  &&  1606.31 &  344.63  &  343.41 &  84.60   &  75.61    \\      
14& $\mathrm{3s3d~^1D_2}$    & 762093  &&  4112.22 &  547.93  &  545.30 &  673.85  &  647.50   \\      
15& $\mathrm{3p3d~^3F^o_2}$  & 928241  &&  876.83  &  485.29  &  485.01 &  $-$125.90 &  $-$126.07  \\  
16& $\mathrm{3p3d~^3F^o_3}$  & 938126  &&  722.00  &  503.29  &  502.30 &  $-$114.13 &  $-$113.47  \\  
17& $\mathrm{3p3d~^1D^o_2}$  & 948513  &&  1621.09 &  408.50  &  408.06 &  $-$176.90 &  $-$176.70  \\  
18& $\mathrm{3p3d~^3F^o_4}$  & 949658  &&  609.19  &  493.11  &  489.87 &  $-$135.80 &  $-$139.09  \\  
19& $\mathrm{3p3d~^3D^o_1}$  & 982868  &&  3141.47 &  349.41  &  348.96 &  $-$114.34 &  $-$120.13  \\  
20& $\mathrm{3p3d~^3P^o_2}$  & 983514  &&  2811.54 &  399.78  &  398.42 &  $-$94.87  &  $-$99.81   \\  
21& $\mathrm{3p3d~^3D^o_3}$  & 994852  &&  3254.23 &  381.16  &  379.36 &  $-$81.04  &  $-$85.84   \\  
22& $\mathrm{3p3d~^3P^o_0}$  & 995889  &&  2550.57 &  467.52  &  461.29 &  $-$58.85  &  $-$78.33   \\  
23& $\mathrm{3p3d~^3P^o_1}$  & 996243  &&  2747.52 &  439.25  &  438.84 &  $-$78.90  &  $-$83.97   \\  
24& $\mathrm{3p3d~^3D^o_2}$  & 996623  &&  2939.80 &  408.62  &  407.89 &  $-$91.96  &  $-$94.83   \\  
25& $\mathrm{3p3d~^1F^o_3}$  & 1062515 &&  4046.73 &  783.97  &  776.35 &  666.63  &  650.87   \\      
26& $\mathrm{3p3d~^1P^o_1}$  & 1074887 &&  3603.75 &  944.60  &  941.74 &  728.85  &  701.36   \\      
27& $\mathrm{3d^2~^3F_2}$    & 1370331 &&  2834.22 &  795.74  &  795.41 &  159.05  &  151.58   \\      
28& $\mathrm{3d^2~^3F_3}$    & 1372035 &&  2752.53 &  776.22  &  775.43 &  148.44  &  139.34   \\      
29& $\mathrm{3d^2~^3F_4}$    & 1374056 &&  2713.16 &  804.78  &  804.19 &  176.26  &  171.07   \\      
30& $\mathrm{3d^2~^1D_2}$    & 1402592 &&  2813.26 &  1079.22 &  1077.95&  458.53  &  442.70   \\      
31& $\mathrm{3d^2~^3P_0}$    &         &&          &          &         &          &           \\      
32& $\mathrm{3d^2~^3P_1}$    &         &&          &          &         &          &           \\      
33& $\mathrm{3d^2~^1G_4}$    & 1407058 &&  2385.95 &  1141.81 &  1139.46&  737.52  &  726.46   \\      
34& $\mathrm{3d^2~^3P_2}$    & 1407773 &&  3038.87 &  932.00  &  931.75 &  195.23  &  187.31   \\      
35& $\mathrm{3d^2~^1S_0}$    & 1487054 &&  2379.42 &  1975.89 &  1975.89&  1143.47 &  1120.02  \\ 
\hline                                                    
\noalign{\smallskip}
\multicolumn{2}{l}{$N_{CSFs}$} & &&  4485&  372043&	430629&	2241061&		5864226  \\
\multicolumn{2}{l}{rms (in cm$^{-1}$)} & &&  2438.25&		653.16&	652.01&		401.16&		392.76  \\
\hline
\hline
\end{tabular}
}
\end{table*}

Table \ref{comp1} shows the total energies from regular {\sc Grasp}2018 calculations 
(\textbf{VV+CV+C+CC RCI int})
for 35 computed states. These energies are computed after CSF base reduction mentioned in previous section. 
Next to this column are the differences of the total energies from \textbf{VV+CV+C+CC RCI} computations, 
which are obtained before a base reduction. The changes in total energies are small, when reduction is applied, 
but the size of the CSF base decreases more than twice. 
In the same Table the energy differences between the \textbf{CV+C+CC~RCI (RSMBPT)}, when 100\% 
of CV, C, CC correlations are included, and regular \textbf{VV+CV+C+CC RCI int} results 
($\Delta E_{\textbf{(CV+C+CC~RCI (RSMBPT))-(VV+CV+C+CC~RCI int)}}$) are given. 
This comparison shows, that the total energies using the RSMBPT method (with reduction column 100\% int 
and without reduction column 100\%) reproduce the regular {\sc Grasp}2018 results.
In case 'int' we have the negligible difference (till 8.0E-06 a.u. or 0.0000000068\%) which could be due to
omitted CV, C and CC correlations with a very small contribution ({\tt 1.0E-11}).

In Table \ref{comp2} the energy levels from regular {\sc Grasp}2018 calculations \textbf{VV+CV+C+CC RCI} and 
from calculations using the RSMBPT method (\textbf{CV+C+CC RCI (RSMBPT)}) are compared. 
The calculations using the RSMBPT method are carried out in five steps, including 
95, 99, 99.5, 99.95 and 100\% of CV, C and CC correlations. 
In the last line of the table the root-mean-square (rms) with results of regular 
{\sc Grasp}2018 calculations (\textbf{VV+CV+C+CC RCI int}) are given. 
By adding the most important $K'$ configurations of CV, C and CC correlations step by step,
the results converge to the results of the regular {\sc Grasp}2018 calculations 
and, in the case '100\% int', are in excellent agreement with them. 
The difference in this case is only up to 2 cm$^{-1}$, which may be due to the omission of CV, C and CC correlations with 
a contribution less than {\tt 1.0E-11} (as mentioned in the discussion of Table \ref{comp1}).
Then CSFs with smaller impact of the CV, C and CC correlations are omitted the energy levels are similar 
to these from the regular computations. For example, if
99\% (case 99\% int) of the CV, C and CC correlations are included in the computations,
the rms with \textbf{VV+CV+C+CC RCI int} results is 68.93 cm$^{-1}$, and the largest difference between the results 
of these calculations is only up to 200 cm$^{-1}$.
Meanwhile, the space of CSFs decreases almost twice comparing to the space in \textbf{VV+CV+C+CC RCI int} computations.

\begin{table*}[!ht]
{\scriptsize
\caption{The total energies (in a.u.) from \textbf{VV+CV+C+CC RCI} calculations and differences (in a.u.) between \textbf{CV+C+CC RCI (RSMBPT)} and \textbf{VV+CV+C+CC RCI int} 
 energies ($\Delta E_{\textbf{(CV+C+CC~RCI (RSMBPT))-(VV+CV+C+CC~RCI int)}}$) for Fe~XV are given when CV, C, and CC correlations are included in the computations.}            
\label{comp1}
\centering
\begin{tabular}{r l r r r r r}
\hline\hline
\multicolumn{1}{c}{\multirow{2}{*}{No.}} & \multicolumn{1}{c}{\multirow{2}{*}{State}} & \multicolumn{2}{c}{\textbf{VV+CV+C+CC RCI}}&& \multicolumn{2}{c}{$\Delta E_{\textbf{(CV+C+CC~RCI (RSMBPT))-(VV+CV+C+CC~RCI int)}}$} \\
\cline{3-4} \cline{6-7}
&&\multicolumn{1}{c}{int}&  \multicolumn{1}{c}{$\Delta E_{\textbf{(VV+CV+C+CC~RCI)-(VV+CV+C+CC~RCI int)}}$} && \multicolumn{1}{c}{100\% int} & \multicolumn{1}{c}{100\%} \\
\hline
\noalign{\smallskip}
 1& $\mathrm{3s^2~^1S_0}$    & -1182.73096327& -0.00003815&&  0.00000009& -0.00001565  \\
 2& $\mathrm{3s3p~^3P^o_0}$  & -1181.66685085& -0.00006259&&  0.00000051& -0.00001094  \\
 3& $\mathrm{3s3p~^3P^o_1}$  & -1181.64029495& -0.00003326&&  0.00000046& -0.00001510  \\
 4& $\mathrm{3s3p~^3P^o_2}$  & -1181.57586867& -0.00002856&&  0.00000019& -0.00001482  \\
 5& $\mathrm{3s3p~^1P^o_1}$  & -1181.12757765& -0.00005702&&  0.00000065& -0.00003104  \\
 6& $\mathrm{3p^2~^3P_0}$    & -1180.20231307& -0.00006452&&  0.00000010& -0.00001675  \\
 7& $\mathrm{3p^2~^1D_2}$    & -1180.17967864& -0.00004235&&  0.00000035& -0.00002700  \\
 8& $\mathrm{3p^2~^3P_1}$    & -1180.15661687& -0.00006994&&  0.00000018& -0.00001747  \\
 9& $\mathrm{3p^2~^3P_2}$    & -1180.07836156& -0.00004763&&  0.00000015& -0.00003757  \\
10& $\mathrm{3p^2~^1S_0}$    & -1179.72222197& -0.00009184&&  0.00000008& -0.00004238  \\
11& $\mathrm{3s3d~^3D_1}$    & -1179.63793828& -0.00010929&&  0.00000111& -0.00002709  \\
12& $\mathrm{3s3d~^3D_2}$    & -1179.63326232& -0.00007751&&  0.00000230& -0.00004417  \\
13& $\mathrm{3s3d~^3D_3}$    & -1179.62581805& -0.00007913&&  0.00000064& -0.00001844  \\
14& $\mathrm{3s3d~^1D_2}$    & -1179.25554176& -0.00015821&&  0.00000409& -0.00008186  \\
15& $\mathrm{3p3d~^3F^o_2}$  & -1178.50215971& -0.00003895&&  0.00000326& -0.00001497  \\
16& $\mathrm{3p3d~^3F^o_3}$  & -1178.45706670& -0.00003518&&  0.00000132& -0.00001109  \\
17& $\mathrm{3p3d~^1D^o_2}$  & -1178.41002605& -0.00003725&&  0.00000086& -0.00001810  \\
18& $\mathrm{3p3d~^3F^o_4}$  & -1178.40462181& -0.00005311&&  0.00000020& -0.00000916  \\
19& $\mathrm{3p3d~^3D^o_1}$  & -1178.25320812& -0.00006454&&  0.00000245& -0.00002416  \\
20& $\mathrm{3p3d~^3P^o_2}$  & -1178.25017600& -0.00006067&&  0.00000054& -0.00002796  \\
21& $\mathrm{3p3d~^3D^o_3}$  & -1178.19845329& -0.00006000&&  0.00000102& -0.00002579  \\
22& $\mathrm{3p3d~^3P^o_0}$  & -1178.19362723& -0.00012693&&  0.00000128& -0.00002239  \\
23& $\mathrm{3p3d~^3P^o_1}$  & -1178.19210564& -0.00006127&&  0.00000281& -0.00003106  \\
24& $\mathrm{3p3d~^3D^o_2}$  & -1178.19043378& -0.00005120&&  0.00000183& -0.00002695  \\
25& $\mathrm{3p3d~^1F^o_3}$  & -1177.88675130& -0.00010997&&  0.00000806& -0.00003279  \\
26& $\mathrm{3p3d~^1P^o_1}$  & -1177.83009684& -0.00016344&&  0.00000423& -0.00006022  \\
27& $\mathrm{3d^2~^3F_2}$    & -1176.48655116& -0.00007218&&  0.00000319& -0.00004057  \\
28& $\mathrm{3d^2~^3F_3}$    & -1176.47883548& -0.00007960&&  0.00000059& -0.00001869  \\
29& $\mathrm{3d^2~^3F_4}$    & -1176.46950039& -0.00006180&&  0.00000028& -0.00003522  \\
30& $\mathrm{3d^2~^1D_2}$    & -1176.33819466& -0.00011029&&  0.00000238& -0.00005909  \\
31& $\mathrm{3d^2~^3P_0}$    & -1176.32567236& -0.00009761&&  0.00000009& -0.00003160  \\
32& $\mathrm{3d^2~^3P_1}$    & -1176.32276134& -0.00009015&&  0.00000057& -0.00002843  \\
33& $\mathrm{3d^2~^1G_4}$    & -1176.31657493& -0.00008852&&  0.00000109& -0.00004105  \\
34& $\mathrm{3d^2~^3P_2}$    & -1176.31578797& -0.00007427&&  0.00000043& -0.00004825  \\
35& $\mathrm{3d^2~^1S_0}$    & -1175.95023666& -0.00014501&&  0.00000009& -0.00006050  \\
\hline
\noalign{\smallskip}
\multicolumn{2}{r}{$N_{CSFs}$} &   2241061&  5864226&&   2166376 & 3033360   \\
\hline
\hline
\end{tabular}
}
\end{table*}

\begin{table*}[!ht]
\setlength{\tabcolsep}{2pt}
{\scriptsize
\caption{The energy levels (in cm$^{-1}$) and differences (in cm$^{-1}$) between \textbf{CV+C+CC~RCI (RSMBPT)} and \textbf{VV+CV+C+CC~RCI int} 
 energies ($\Delta E_{\textbf{(CV+C+CC~RCI (RSMBPT))-(VV+CV+C+CC~RCI int)}}$) for Fe~XV are given when CV, C and CC correlations are included in the computations.}
\label{comp2}
\centering
\begin{tabular}{r l r r r r r r r r r r r r r}
\hline\hline
\multicolumn{1}{c}{\multirow{2}{*}{No.}} & \multicolumn{1}{c}{\multirow{2}{*}{State}} 
 & \multicolumn{2}{c}{\textbf{VV+CV+C+CC RCI}}&& \multicolumn{10}{c}{$\Delta E_{\textbf{(CV+C+CC~RCI (RSMBPT))-(VV+CV+C+CC~RCI int)}}$} \\
\cline{3-4} \cline{6-15}
&&& \multicolumn{1}{c}{int} && 
\multicolumn{1}{c}{95\%} & \multicolumn{1}{c}{95\% int} & 
\multicolumn{1}{c}{99\%} & \multicolumn{1}{c}{99\% int} & 
\multicolumn{1}{c}{99.5\%} & \multicolumn{1}{c}{99.5\% int} &
\multicolumn{1}{c}{99.95\%} & \multicolumn{1}{c}{99.95\% int} &
\multicolumn{1}{c}{100\%}  & \multicolumn{1}{c}{100\% int} \\
\hline
\noalign{\smallskip}
 1& $\mathrm{3s^2~^1S_0}$    &      0.00  &      0.00  &&         &        &         &        &        &         &        &        &       &      \\
 2& $\mathrm{3s3p~^3P^o_0}$  & 233540.32  & 233545.68  && 269.24  & 268.63 &  98.71  & 97.70  & 86.10  &  85.17  & 28.91  & 28.03  & 1.04  & 0.09 \\
 3& $\mathrm{3s3p~^3P^o_1}$  & 239375.10  & 239374.03  && 26.84   & 26.50  &  -7.13  & -7.51  & 3.96   &  3.70   & 1.67   & 1.67   & 0.12  & 0.08 \\
 4& $\mathrm{3s3p~^3P^o_2}$  & 253516.07  & 253513.96  && 107.85  & 107.63 &  30.50  & 30.16  & 19.82  &  19.50  & 6.30   & 6.19   & 0.18  & 0.02 \\
 5& $\mathrm{3s3p~^1P^o_1}$  & 351898.33  & 351902.47  && -27.93  & -25.84 &  -28.78 & -26.15 & -15.29 &  -12.44 & -3.59  & -0.11  & -3.38 & 0.12 \\
 6& $\mathrm{3p^2~^3P_0}$    & 554968.78  & 554974.57  && 95.62   & 95.68  &  22.36  & 22.31  & 24.60  &  24.59  & 11.46  & 11.55  & -0.24 & 0.00 \\
 7& $\mathrm{3p^2~^1D_2}$    & 559941.33  & 559942.25  && -20.74  & -19.25 &  -27.95 & -26.27 & -31.06 &  -29.26 & -11.44 & -9.34  & -2.49 & 0.06 \\
 8& $\mathrm{3p^2~^3P_1}$    & 564996.75  & 565003.73  && 138.27  & 138.53 &  42.76  & 42.84  & 18.54  &  18.66  & 4.03   & 4.28   & -0.40 & 0.02 \\
 9& $\mathrm{3p^2~^3P_2}$    & 582176.70  & 582178.78  && -66.36  & -62.91 &  -73.46 & -69.68 & -58.10 &  -54.15 & -15.52 & -11.27 & -4.81 & 0.02 \\
10& $\mathrm{3p^2~^1S_0}$    & 660330.60  & 660342.39  && 65.22   & 69.98  &  7.13   & 12.31  & 1.12   &  6.44   & 0.21   & 5.72   & -5.87 & 0.00 \\
11& $\mathrm{3s3d~^3D_1}$    & 678824.91  & 678840.52  && 67.36   & 68.47  &  27.54  & 29.14  & -1.80  &  -0.05  & -5.67  & -3.55  & -2.51 & 0.22 \\
12& $\mathrm{3s3d~^3D_2}$    & 679858.14  & 679866.77  && -149.64 & -146.78&  -64.98 & -61.02 & -49.64 &  -45.30 & -23.87 & -18.62 & -6.26 & 0.49 \\
13& $\mathrm{3s3d~^3D_3}$    & 681491.61  & 681500.60  && 130.96  & 130.78 &  36.85  & 36.83  & 23.13  &  23.19  & -9.37  & -8.87  & -0.61 & 0.12 \\
14& $\mathrm{3s3d~^1D_2}$    & 762740.50  & 762766.85  && -111.94 & -105.07&  -65.59 & -55.96 & -52.66 &  -42.20 & -27.10 & -14.63 & -14.53& 0.88 \\
15& $\mathrm{3p3d~^3F^o_2}$  & 928114.93  & 928115.10  && 113.06  & 112.34 &  56.94  & 56.38  & 39.81  &  39.68  & 4.38   & 4.85   & 0.15  & 0.70 \\
16& $\mathrm{3p3d~^3F^o_3}$  & 938012.53  & 938011.87  && 326.85  & 325.85 &  109.11 & 108.03 & 67.58  &  66.59  & 13.83  & 13.08  & 1.01  & 0.27 \\
17& $\mathrm{3p3d~^1D^o_2}$  & 948336.30  & 948336.10  && 70.58   & 70.43  &  15.00  & 14.99  & 3.96   &  4.13   & -3.70  & -3.08  & -0.53 & 0.17 \\
18& $\mathrm{3p3d~^3F^o_4}$  & 949518.91  & 949522.20  && 491.20  & 489.86 &  184.61 & 183.15 & 109.42 &  107.98 & 21.67  & 20.32  & 1.42  & 0.02 \\
19& $\mathrm{3p3d~^3D^o_1}$  & 982747.87  & 982753.66  && 167.53  & 167.81 &  35.77  & 36.50  & 28.10  &  29.35  & 0.41   & 2.38   & -1.87 & 0.52 \\
20& $\mathrm{3p3d~^3P^o_2}$  & 983414.19  & 983419.13  && 93.92   & 94.79  &  26.46  & 27.94  & 10.51  &  12.16  & -3.24  & -0.81  & -2.70 & 0.11 \\
21& $\mathrm{3p3d~^3D^o_3}$  & 994766.16  & 994770.96  && 227.51  & 228.31 &  80.34  & 81.70  & 51.84  &  53.48  & 3.12   & 5.37   & -2.23 & 0.20 \\
22& $\mathrm{3p3d~^3P^o_0}$  & 995810.67  & 995830.15  && 357.05  & 358.06 &  190.21 & 191.71 & 105.13 &  106.82 & 27.32  & 29.11  & -1.47 & 0.27 \\
23& $\mathrm{3p3d~^3P^o_1}$  & 996159.03  & 996164.10  && 83.70   & 84.84  &  -2.46  & -0.44  & -5.12  &  -2.19  & -6.24  & -2.48  & -3.38 & 0.60 \\
24& $\mathrm{3p3d~^3D^o_2}$  & 996528.17  & 996531.04  && 10.93   & 11.81  &  -4.60  & -3.05  & -6.93  &  -5.06  & -7.33  & -4.61  & -2.48 & 0.38 \\
25& $\mathrm{3p3d~^1F^o_3}$  & 1063165.87 & 1063181.63 && 250.48  & 252.57 &  93.45  & 96.76  & 61.00  &  64.93  & 6.67   & 11.86  & -3.75 & 1.76 \\
26& $\mathrm{3p3d~^1P^o_1}$  & 1075588.36 & 1075615.85 && 114.33  & 117.36 &  36.38  & 42.97  & 16.88  &  24.45  & -8.23  & 1.69   & -9.78 & 0.91 \\
27& $\mathrm{3d^2~^3F_2}$    & 1370482.58 & 1370490.05 && -144.36 & -142.48&  -70.01 & -66.81 & -52.54 &  -48.78 & -23.73 & -19.12 & -5.47 & 0.68 \\
28& $\mathrm{3d^2~^3F_3}$    & 1372174.34 & 1372183.44 && 217.40  & 217.20 &  52.82  & 52.80  & 25.63  &  25.74  & -5.82  & -5.50  & -0.66 & 0.11 \\
29& $\mathrm{3d^2~^3F_4}$    & 1374227.07 & 1374232.26 && 121.04  & 122.80 &  7.66   & 10.51  & -19.51 &  -16.53 & -13.99 & -10.42 & -4.30 & 0.04 \\
30& $\mathrm{3d^2~^1D_2}$    & 1403034.70 & 1403050.53 && -209.47 & -205.43&  -90.09 & -83.82 & -68.86 &  -62.04 & -30.38 & -22.18 & -9.53 & 0.51 \\
31& $\mathrm{3d^2~^3P_0}$    & 1405785.81 & 1405798.86 && 143.08  & 145.54 &  34.99  & 37.82  & 22.03  &  25.00  & -0.39  & 2.85   & -3.50 & 0.00 \\
32& $\mathrm{3d^2~^3P_1}$    & 1406426.34 & 1406437.76 && 171.29  & 173.07 &  40.32  & 42.53  & 8.35   &  10.72  & -9.14  & -6.58  & -2.81 & 0.10 \\
33& $\mathrm{3d^2~^1G_4}$    & 1407784.46 & 1407795.52 && -15.12  & -10.90 &  38.56  & 40.54  & 10.65  &  12.84  & -8.82  & -5.68  & -5.58 & 0.22 \\
34& $\mathrm{3d^2~^3P_2}$    & 1407960.31 & 1407968.23 && -13.82  & -13.14 &  -91.34 & -85.61 & -73.08 &  -67.10 & -30.13 & -23.50 & -7.15 & 0.08 \\
35& $\mathrm{3d^2~^1S_0}$    & 1488174.02 & 1488197.47 && 113.10  & 120.88 &  27.43  & 35.89  & 15.96  &  24.61  & -6.97  & 2.21   & -9.84 & 0.00 \\
\hline                                                    
\noalign{\smallskip}
\multicolumn{2}{l}{$N_{CSFs}$} &    5864226&  2241061&& 1030900&	717523& 1777303&	1227367& 2033820&	1413852& 2607682&	1840752& 3033360&  2166376  \\
\multicolumn{2}{l}{rms (in cm$^{-1}$)} & &  && 174.40&	174.39& 69.49&	68.93& 45.63&	44.70& 14.69&	12.35& 4.91&  0.46  \\
\hline
\hline
\end{tabular}
}
\end{table*}

\subsection{Calculation of E1 transition properties}
\label{sec:Calculation_E1}
\begin{figure}
\centering
\includegraphics[scale=0.4]{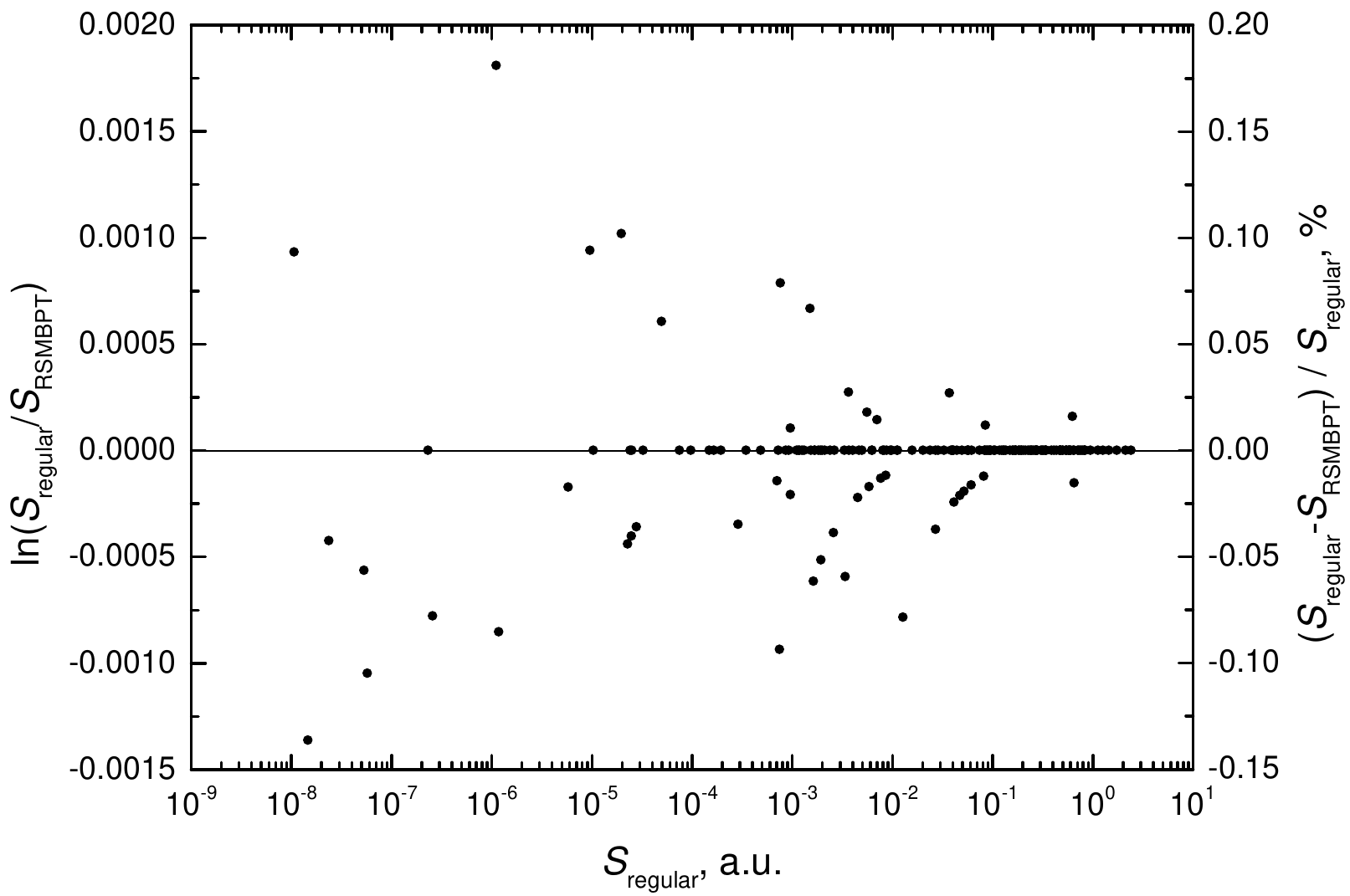}
\caption{\label{1_graph} Comparison of all line strengths 
from the regular {\sc Grasp}2018 calculations (\textbf{VV+CV+C+CC RCI int}) 
with the results from 
the \textbf{CV+C+CC~RCI (RSMBPT) int} strategy, when 100\% of the CV, C, CC correlations are included.
The line strengths are compared in the Babushkin gauge.}
\end{figure}

\begin{figure}
\centering
\includegraphics[scale=0.4]{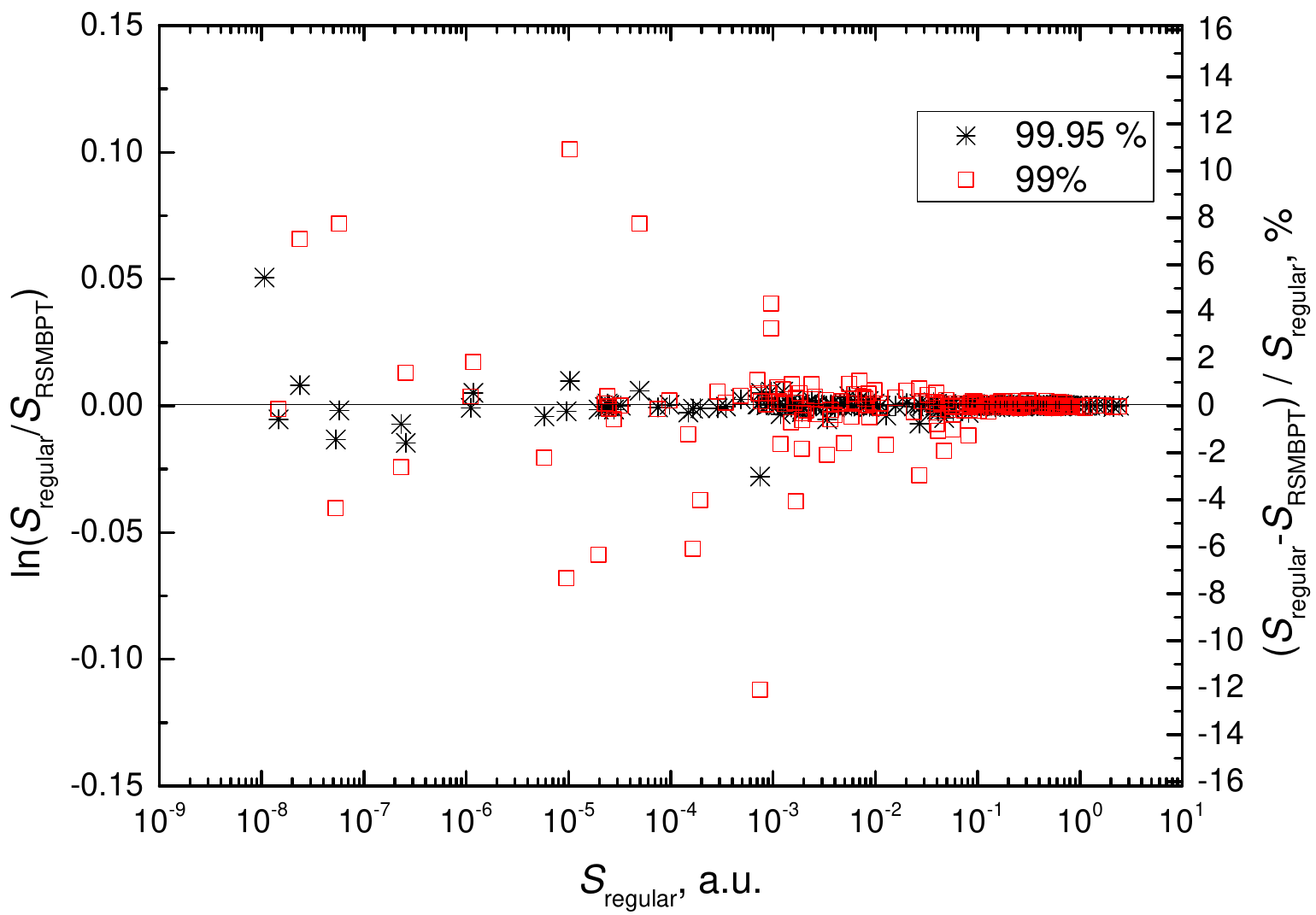}
\caption{\label{2_graph}  Comparison of all line strengths 
from the regular {\sc Grasp}2018 calculations (\textbf{VV+CV+C+CC RCI int}) 
with the results from 
the \textbf{CV+C+CC~RCI (RSMBPT) int} strategy, when when 99.95 and 99\% of the CV, C, CC correlations are included.
The line strengths are compared in the Babushkin gauge.} 
\end{figure}

\begin{figure}
\centering
\includegraphics[scale=0.4]{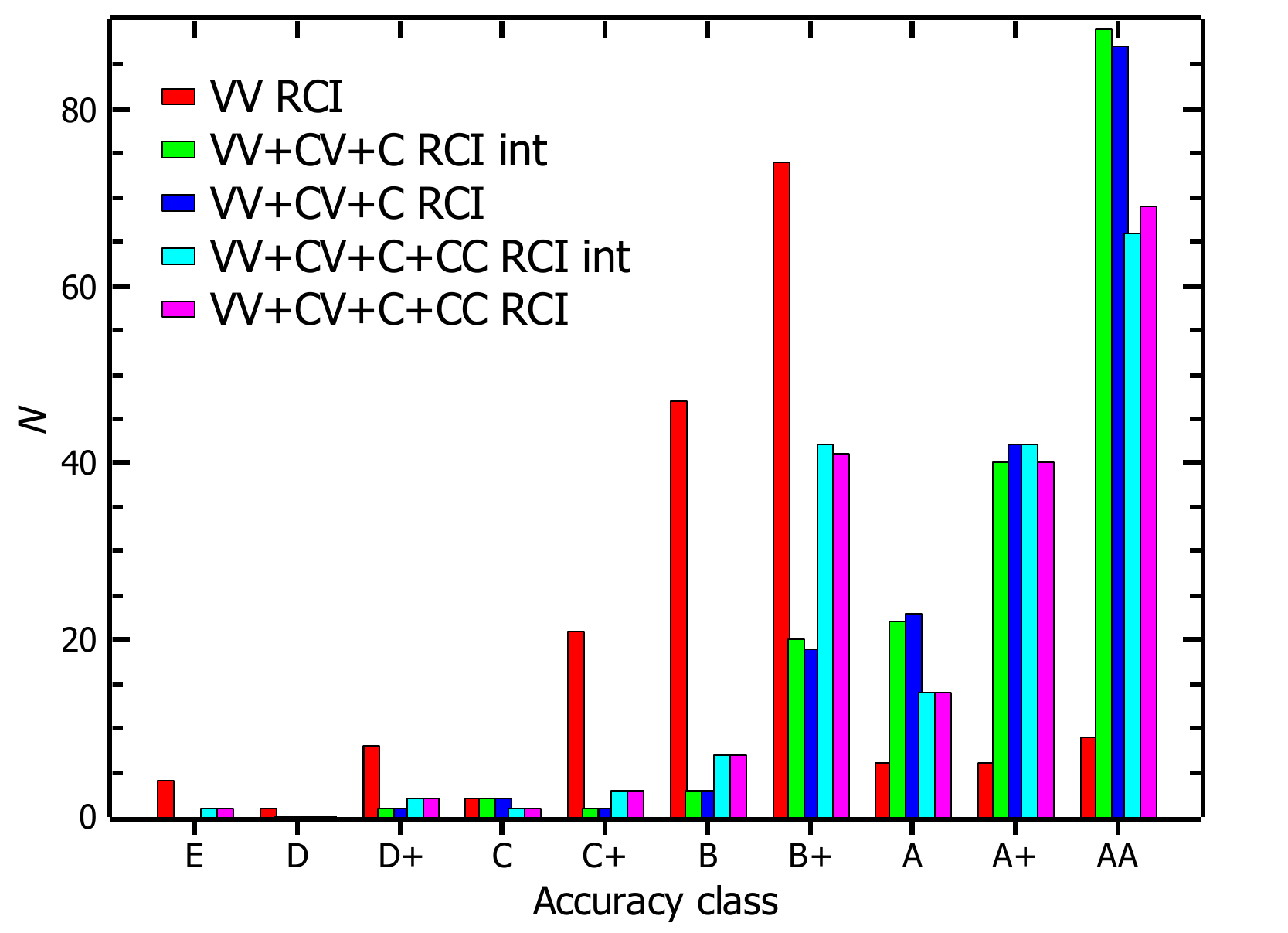}
\caption{\label{statistika_GRASP_regular} Distribution of E1 transitions over the accuracy classes 
according to the computational schemes.}
\end{figure}

\begin{figure}
\centering
\includegraphics[scale=0.4]{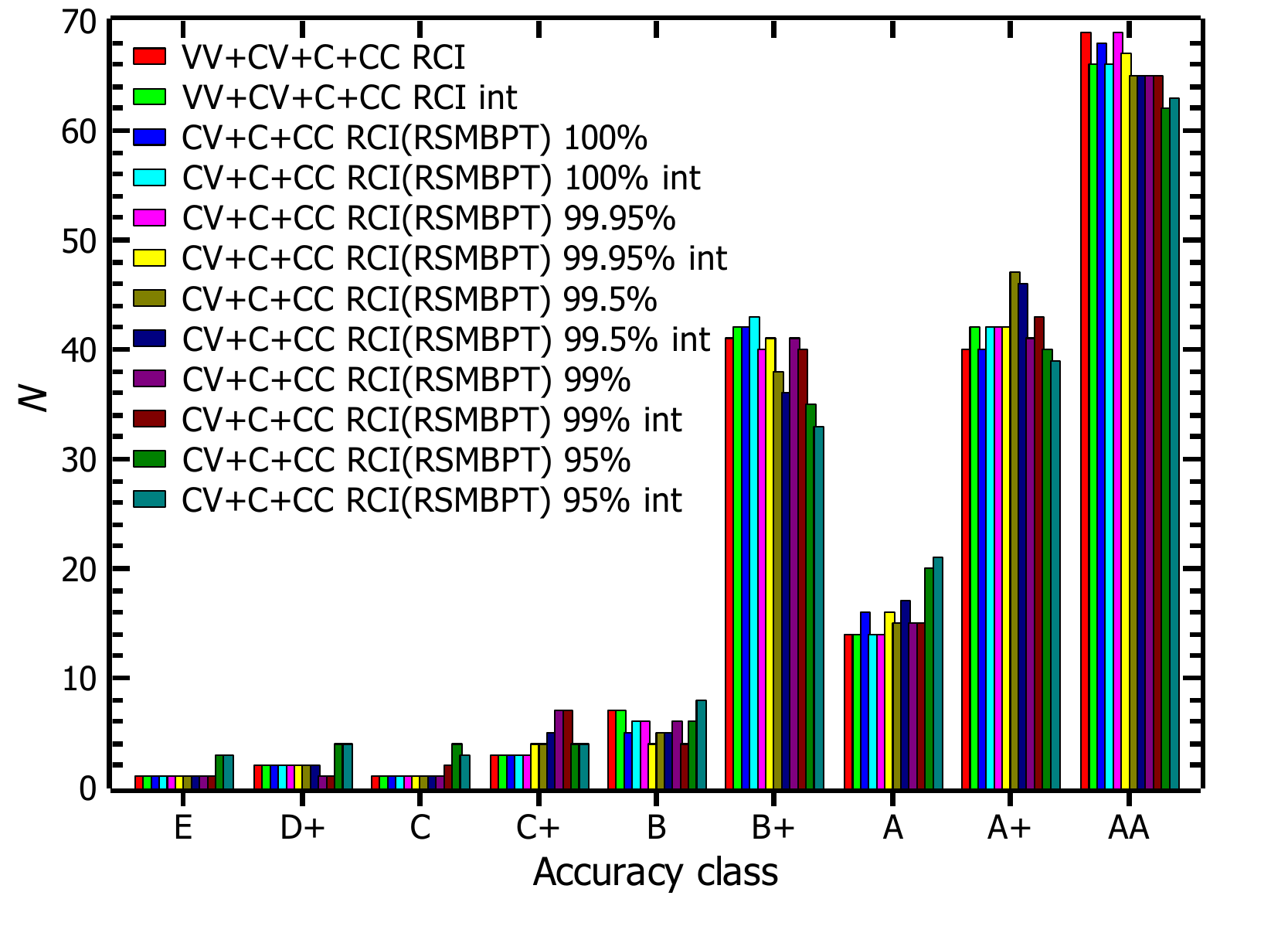}
\caption{\label{statistika_PT2}  Distribution of E1 transitions over the accuracy classes 
according to the computational schemes. The results from regular {\sc Grasp}2018 calculations 
are compared with the results using RSMBPT when a different amount of CV, C, CC correlations are included. }
\end{figure}

We computed the transition properties of E1 transitions
between the above mentioned levels. The calculations are done in a regular way by including different types of correlations and using 
the RSMBPT method. Below we present a comparison of the line strengths using different computational schemes 
(Table \ref{com_line_st_schemes} and Figs. \ref{1_graph}, \ref{2_graph}).
We also give the statistics for all computed transitions (Figs. \ref{statistika_GRASP_regular} and \ref{statistika_PT2}). 
The uncertainties of the line strengths obtained in this work are estimated based on the quantitative and qualitative evaluation
(QQE) method described in Refs. \cite{Ce_IV,Pr_IV}.

Table \ref{com_line_st_schemes} shows the comparison of the line strengths, cancellation factors (CF) \cite{Cowan}, 
the $G_{S=0}$ parameters \cite{Ce_IV,Pr_IV}, and the estimated accuracy for few strongest 
and for few weaker E1 transitions using different computational schemes.
The table also includes lines for each transition marked with 'NIST' as these data are critically evaluated by the NIST \cite{NIST_ASD}. 
These lines contain the observed wavelength and line strength with the critically evaluated accuracy.
As seen from the table, using the regular computational scheme, the CV and C correlations have the largest impact on the line strengths, 
adding the CC correlations the line strengths change a little. By using the RSMBPT method it is seen, that by including 
the most important $K'$ configurations of CV, C and CC correlations the line strengths agree with the regular {\sc Grasp}2018 calculations,
when CV, C and CC correlations are included. 
Excluding the CV, C, CC correlations with the smallest impact (cases 99.95, 99.5, 99 and 95\% in Table \ref{com_line_st_schemes}), 
we see that the line strength almost does not change comparing to the results when all these correlations are included.
From Table \ref{com_line_st_schemes} we also see that the CFs in both (Babushkin and Coulomb) gauges change 
when a new group of correlations is added to the regular {\sc Grasp}2018 calculations.
Using the RSMBPT method, the values of CF are stable when the most important configurations 
of CV, C, and CC correlations in different amount are included and 
the configurations of CV, C, and CC correlations with smallest impact are neglected.
The change in the values of CFs indicates the importance of added correlations in the calculations.
By using regular method for inclusion the new group of correlations we can have such cases (as the last transition 
in Table \ref{com_line_st_schemes}) when values of line strength from two computational strategies disagree more 
than the accuracy class was assigned for transition. 
This situation was observed for some transitions assigned with the best accuracy classes.
For example, the line strength of the $\mathrm{3s\,3p~^3P^o_1}	$ -- $\mathrm{3p^2~^1D_2}	$ transition 
in both (\textbf{VV+CV+C~RCI int} and \textbf{VV+CV+C+CC~RCI int}) strategies is assigned to the AA accuracy class.
In the regular {\sc Grasp}2018 calculations in the \textbf{VV+CV+C~RCI int} strategy the line strength in the Babushkin gauge is 9.14394E-02 a.u., 
meanwhile when CC correlations were added (\textbf{VV+CV+C+CC~RCI int} strategy) the line strength in the Babushkin gauge is equal 8.99187E-02 a.u.
Whereas using the RSMBPT method when the CV, C, and CC correlations are included 
the line strength in the Babushkin gauge is close to 8.9E-02 a.u. value.
Therefore, using the RSMBPT method to include the CV, C, and CC correlations is more efficient than the regular method 
because it allows to estimate the impact of the correlations and include the most important from all correlations types rather 
than the whole group of correlations.
The use of the RSMBPT method for the inclusion of CV, C, and CC correlations also significantly reduces CSF space
and therefore for more complex atom or ion allows include all types of correlations 
which is not always possible in a regular {\sc Grasp}2018 calculations.
Thus, the QQE method provides a more accurate estimate of the errors in the case when the RSMBPT method is used.

The line strengths from present computations are similar to those provided by NIST.. 
Most of the line strengths agree better  
when only VV correlations are included (\textbf{VV~RCI} strategy,), 
instead when all types of the correlations are taken into account.
 Also it should be mentioned that the line strengths given in the NIST
are assigned with worse accuracy class than in present computations.

Figure \ref{1_graph} presents the comparison of all line strengths 
from the regular {\sc Grasp}2018 calculations (\textbf{VV+CV+C+CC RCI int}) 
with the results from 
the \textbf{CV+C+CC~RCI (RSMBPT) int} strategy, when 100\% of the CV, C, CC correlations are included.
The line strengths are compared in the Babushkin gauge.
It is seen that the line strengths obtained using the RSMBPT method reproduce the results of the regular calculations. 
The line strengths between the two calculations fully agree for most of the transitions, 
and differ only to within 0.18\% for some of the weakest transitions.
Figure \ref{2_graph} presents the comparison of all line strengths
from the regular {\sc Grasp}2018 calculations (\textbf{VV+CV+C+CC RCI int}) 
with the results from 
the \textbf{CV+C+CC~RCI (RSMBPT) int} strategy, when 99.95\% and 99\% of the CV, C, CC correlations are included.
From the figure weakest transition ($S$=1.07E-08) is excluded in case 99\%, 
as the natural logarithm ratio for this transition is equal 0.36. 
The line strengths between the regular {\sc Grasp}2018 calculations 
and calculations using RSMBPT method (cases 99.95\% and 99\%) agree well for most of the transitions.
In the case 99.95\% the discrepancy between two calculations is only up to 1.5\%, except for two transitions where 
it reaches 2.8\% and 5.2\%.
In the case 99\% the strongest transitions agree very well between two calculations, the discrepancy for 
weaker transitions reach up to 12\% for some lines.

Figure \ref{statistika_GRASP_regular} shows the statistical distribution of all computed E1 transitions 
by accuracy classes according to the computational schemes using the regular {\sc Grasp}2018 method. 
It can be seen that the inclusion of the CV, C correlations 
improves the agreement between the Babushkin and Coulomb gauges. By adding the CC correlations
the number of transitions assigned with AA accuracy class reduced a little, and in B+ accuracy class it increases.

Figure \ref{statistika_PT2} shows the statistical distributions of all computed E1 transitions 
by accuracy class when CV, C, and CC correlations with different percentage of correlations are included  
using the RSMBPT method and using the regular {\sc Grasp}2018 computational scheme. 
From the figure it is seen that the transition distributions by accuracy classes
are similar in both these methods, and remains similar when configurations of CV, C, and CC correlations with smallest impact are neglected.

{\scriptsize
\begin{longtable}{l l r r r r r l}
\caption{\label{com_line_st_schemes} Comparison of computed wavelengths ($\lambda$ in \AA), line strengths ($S$ in a.u.), 
cancellation factors, and the $G_{S=0}$ parameters using different strategies. 
$S_B$ is the line strength in the Babushkin gauge, $S_C$ is the line strength in the Coulomb gauge. 
The name of the computational scheme \textbf{CV+C+CC~RCI(RSMBPT)} is marked as \textbf{RCI~(RSMBPT)} in the table.}\\
\hline\hline
\multicolumn{1}{c}{Strategy} & \multicolumn{1}{c}{$\lambda$} & \multicolumn{1}{c}{$S_B$} & \multicolumn{1}{c}{$S_C$} & \multicolumn{1}{c}{CF$_B$} & \multicolumn{1}{c}{CF$_C$} & \multicolumn{1}{c}{$G_{S=0}$} & \multicolumn{1}{c}{Acc.}\\
\hline
\noalign{\smallskip}
\endfirsthead
\caption{Continued.}\\
\hline\hline
\multicolumn{1}{c}{Strategy} & \multicolumn{1}{c}{$\lambda$} & \multicolumn{1}{c}{$S_B$} & \multicolumn{1}{c}{$S_C$} & \multicolumn{1}{c}{CF$_B$} & \multicolumn{1}{c}{CF$_C$} & \multicolumn{1}{c}{$G_{S=0}$} & \multicolumn{1}{c}{Acc.}\\
\hline
\noalign{\smallskip}
\endhead
\noalign{\smallskip}
\hline
\hline
\endfoot
\noalign{\smallskip}
\multicolumn{8}{c}{$\mathrm{3s\,3p~^1P^o_1}	$ -- $\mathrm{3s\,3d~^1D_2}	$} \\
\textbf{VV~RCI}	                  &243.09 &1.50687E+00& 1.58586E+00& 7.07E-01& 7.98E-01&   5.60659E+01& B+ \\                      
\textbf{VV+CV+C~RCI int}          &243.59& 1.43765E+00& 1.43571E+00& 6.59E-01& 3.83E-01&$-$2.09378E+03& AA \\ 
\textbf{VV+CV+C~RCI}              &243.60& 1.43765E+00& 1.43577E+00& 6.59E-01& 3.83E-01&$-$2.15614E+03& AA \\                                 
\textbf{VV+CV+C+CC~RCI int}       &243.39& 1.44376E+00& 1.44403E+00& 6.60E-01& 3.83E-01&   1.48600E+04& AA \\              
\textbf{VV+CV+C+CC~RCI}	          &243.40& 1.44367E+00& 1.44383E+00& 6.59E-01& 3.82E-01&   2.59365E+04& AA \\           
\textbf{RCI~(RSMBPT) 100\%}       &243.40& 1.44367E+00& 1.44370E+00& 6.60E-01& 3.82E-01&   1.44867E+05& AA \\      
\textbf{RCI~(RSMBPT) 100\% int}	  &243.39& 1.44376E+00& 1.44402E+00& 6.60E-01& 3.83E-01&   1.55237E+04& AA \\  
\textbf{RCI~(RSMBPT) 99.95\%}	    &243.40& 1.44371E+00& 1.44386E+00& 6.60E-01& 3.83E-01&   2.62284E+04& AA \\    
\textbf{RCI~(RSMBPT) 99.95\% int} &243.40& 1.44380E+00& 1.44420E+00& 6.60E-01& 3.83E-01&   1.00242E+04& AA \\
\textbf{RCI~(RSMBPT) 99.5\%}      &243.41& 1.44379E+00& 1.44434E+00& 6.60E-01& 3.83E-01&   7.46809E+03& AA \\     
\textbf{RCI~(RSMBPT) 99.5\% int}  &243.41& 1.44389E+00& 1.44471E+00& 6.60E-01& 3.84E-01&   4.96748E+03& AA \\ 
\textbf{RCI~(RSMBPT) 99\%}	      &243.41& 1.44392E+00& 1.44493E+00& 6.60E-01& 3.84E-01&   4.04749E+03& AA \\       
\textbf{RCI~(RSMBPT) 99\% int}	  &243.41& 1.44401E+00& 1.44530E+00& 6.60E-01& 3.85E-01&   3.15843E+03& AA \\   
\textbf{RCI~(RSMBPT) 95\%}	      &243.44& 1.44460E+00& 1.44562E+00& 6.60E-01& 3.88E-01&   4.03134E+03& AA \\       
\textbf{RCI~(RSMBPT) 95\% int}	  &243.44& 1.44470E+00& 1.44599E+00& 6.60E-01& 3.89E-01&   3.16669E+03& AA \\ 
NIST \cite{NIST_ASD}            &243.794& 1.5E+00& &&&& D\\
\\
\multicolumn{8}{c}{$\mathrm{3p^2~^3P_2}	$ -- $\mathrm{3p\,3d~^3D^o_3}	$} \\
\textbf{VV~RCI}	                  & 241.04& 1.20534E+00& 1.28606E+00& 5.05E$-$01& 5.42E$-$01& 4.43456E+01 & B+ \\ 
\textbf{VV+CV+C~RCI int}          & 241.98& 1.11498E+00& 1.10652E+00& 4.57E$-$01& 2.59E$-$01&$-$3.70647E+02& AA \\
\textbf{VV+CV+C~RCI}              & 241.98& 1.11499E+00& 1.10653E+00& 4.57E$-$01& 2.59E$-$01&$-$3.70698E+02& AA \\
\textbf{VV+CV+C+CC~RCI int}       & 242.37& 1.12568E+00& 1.12684E+00& 4.60E$-$01& 2.60E$-$01& 2.76724E+03 & AA \\ 
\textbf{VV+CV+C+CC~RCI}	          & 242.37& 1.12551E+00& 1.12672E+00& 4.60E$-$01& 2.60E$-$01& 2.64286E+03 & AA \\ 
\textbf{RCI~(RSMBPT) 100\%}       & 242.37& 1.12557E+00& 1.12677E+00& 4.60E$-$01& 2.60E$-$01& 2.66400E+03 & AA \\ 
\textbf{RCI~(RSMBPT) 100\% int}	  & 242.37& 1.12569E+00& 1.12684E+00& 4.60E$-$01& 2.60E$-$01& 2.76393E+03 & AA \\
\textbf{RCI~(RSMBPT) 99.95\%}     & 242.36& 1.12575E+00& 1.12637E+00& 4.60E$-$01& 2.61E$-$01& 5.16207E+03 & AA \\
\textbf{RCI~(RSMBPT) 99.95\% int} & 242.36& 1.12586E+00& 1.12647E+00& 4.60E$-$01& 2.61E$-$01& 5.21141E+03 & AA \\
\textbf{RCI~(RSMBPT) 99.5\%}      & 242.31& 1.12621E+00& 1.12583E+00& 4.60E$-$01& 2.63E$-$01&$-$8.45922E+03& AA \\
\textbf{RCI~(RSMBPT) 99.5\% int}  & 242.31& 1.12631E+00& 1.12595E+00& 4.60E$-$01& 2.63E$-$01&$-$8.81998E+03& AA \\
\textbf{RCI~(RSMBPT) 99\%}	      & 242.28& 1.12673E+00& 1.12562E+00& 4.60E$-$01& 2.64E$-$01&$-$2.87810E+03& AA \\
\textbf{RCI~(RSMBPT) 99\% int}	  & 242.28& 1.12684E+00& 1.12575E+00& 4.60E$-$01& 2.64E$-$01&$-$2.94103E+03& AA \\
\textbf{RCI~(RSMBPT) 95\%}	      & 242.20& 1.12779E+00& 1.12878E+00& 4.61E$-$01& 2.67E$-$01& 3.22883E+03 & AA \\ 
\textbf{RCI~(RSMBPT) 95\% int}	  & 242.20& 1.12789E+00& 1.12891E+00& 4.61E$-$01& 2.67E$-$01& 3.13652E+03 & AA \\
NIST \cite{NIST_ASD}            &242.100 & 1.1E+00& &&&& D\\
\\
\multicolumn{8}{c}{$\mathrm{3s\,3d~^1D_2}	$ -- $\mathrm{3p\,3d~^1F^o_3}	$} \\
\textbf{VV~RCI}	                  & 332.94& 2.21030E+00& 2.22678E+00& 7.54E-01& 3.54E-01& 3.81508E+02& AA  \\
\textbf{VV+CV+C~RCI int}          & 332.60& 2.11114E+00& 2.12236E+00& 7.04E-01& 1.93E-01& 5.34174E+02& AA  \\
\textbf{VV+CV+C~RCI}              & 332.61& 2.11102E+00& 2.12186E+00& 7.04E-01& 1.93E-01& 5.53027E+02& AA  \\
\textbf{VV+CV+C+CC~RCI int}       & 332.87& 2.12945E+00& 2.20008E+00& 7.10E-01& 2.00E-01& 8.73887E+01& B+  \\
\textbf{VV+CV+C+CC~RCI}	          & 332.86& 2.12925E+00& 2.19992E+00& 7.10E-01& 2.00E-01& 8.73377E+01& B+  \\
\textbf{RCI~(RSMBPT) 100\%}       & 332.86& 2.12938E+00& 2.20043E+00& 7.10E-01& 2.00E-01& 8.68847E+01& B+  \\
\textbf{RCI~(RSMBPT) 100\% int}	  & 332.87& 2.12948E+00& 2.20021E+00& 7.10E-01& 2.00E-01& 8.72685E+01& B+  \\
\textbf{RCI~(RSMBPT) 99.95\%}     & 332.84& 2.12948E+00& 2.19936E+00& 7.10E-01& 2.00E-01& 8.83069E+01& B+  \\
\textbf{RCI~(RSMBPT) 99.95\% int} & 332.84& 2.12957E+00& 2.19922E+00& 7.10E-01& 2.00E-01& 8.85934E+01& B+  \\
\textbf{RCI~(RSMBPT) 99.5\%}      & 332.75& 2.12964E+00& 2.19978E+00& 7.11E-01& 2.01E-01& 8.79876E+01& B+  \\
\textbf{RCI~(RSMBPT) 99.5\% int}  & 332.75& 2.12973E+00& 2.19970E+00& 7.11E-01& 2.01E-01& 8.82038E+01& B+  \\
\textbf{RCI~(RSMBPT) 99\%}	      & 332.70& 2.12982E+00& 2.19981E+00& 7.11E-01& 2.01E-01& 8.81840E+01& B+  \\
\textbf{RCI~(RSMBPT) 99\% int}	  & 332.70& 2.12991E+00& 2.19976E+00& 7.11E-01& 2.01E-01& 8.83611E+01& B+  \\
\textbf{RCI~(RSMBPT) 95\%}	      & 332.47& 2.13084E+00& 2.19738E+00& 7.11E-01& 2.04E-01& 9.26952E+01& B+  \\
\textbf{RCI~(RSMBPT) 95\% int}	  & 332.48& 2.13093E+00& 2.19740E+00& 7.11E-01& 2.04E-01& 9.27792E+01& B+  \\
NIST \cite{NIST_ASD}            &332.854 & 2.3E+00& &&&& D\\
\\
\multicolumn{8}{c}{$\mathrm{3s\,3d~^3D_3}	$ -- $\mathrm{3p\,3d~^3F^o_4}	$} \\
\textbf{VV~RCI}	                  & 374.19& 1.30935E+00& 1.26293E+00& 8.84E-01& 4.79E-01&$-$7.76641E+01 & B+  \\
\textbf{VV+CV+C~RCI int}          & 372.59& 1.26123E+00& 1.28633E+00& 8.20E-01& 2.31E-01&   1.44239E+02 & A+  \\
\textbf{VV+CV+C~RCI}              & 372.59& 1.26107E+00& 1.28489E+00& 8.20E-01& 2.30E-01&   1.51858E+02 & A+  \\
\textbf{VV+CV+C+CC~RCI int}       & 373.10& 1.26785E+00& 1.31118E+00& 8.26E-01& 2.36E-01&   8.48785E+01 & B+  \\
\textbf{VV+CV+C+CC~RCI}	          & 373.10& 1.26760E+00& 1.31038E+00& 8.26E-01& 2.35E-01&   8.59229E+01 & B+  \\
\textbf{RCI~(RSMBPT) 100\%}       & 373.10& 1.26782E+00& 1.31121E+00& 8.26E-01& 2.36E-01&   8.47539E+01 & B+  \\
\textbf{RCI~(RSMBPT) 100\% int}	  & 373.10& 1.26785E+00& 1.31117E+00& 8.26E-01& 2.36E-01&   8.49020E+01 & B+  \\
\textbf{RCI~(RSMBPT) 99.95\%}     & 373.06& 1.26783E+00& 1.31075E+00& 8.26E-01& 2.36E-01&   8.56675E+01 & B+  \\
\textbf{RCI~(RSMBPT) 99.95\% int} & 373.06& 1.26786E+00& 1.31071E+00& 8.26E-01& 2.36E-01&   8.57989E+01 & B+  \\
\textbf{RCI~(RSMBPT) 99.5\%}      & 372.98& 1.26787E+00& 1.31123E+00& 8.26E-01& 2.37E-01&   8.48162E+01 & B+  \\
\textbf{RCI~(RSMBPT) 99.5\% int}  & 372.99& 1.26790E+00& 1.31124E+00& 8.27E-01& 2.37E-01&   8.48627E+01 & B+  \\
\textbf{RCI~(RSMBPT) 99\%}	      & 372.90& 1.26795E+00& 1.31118E+00& 8.27E-01& 2.38E-01&   8.50709E+01 & B+  \\
\textbf{RCI~(RSMBPT) 99\% int}	  & 372.90& 1.26798E+00& 1.31119E+00& 8.27E-01& 2.38E-01&   8.51050E+01 & B+  \\
\textbf{RCI~(RSMBPT) 95\%}	      & 372.60& 1.26875E+00& 1.30879E+00& 8.28E-01& 2.45E-01&   9.17442E+01 & B+  \\
\textbf{RCI~(RSMBPT) 95\% int}	  & 372.61& 1.26877E+00& 1.30885E+00& 8.28E-01& 2.45E-01&   9.16625E+01 & B+  \\
NIST \cite{NIST_ASD}            &372.798 & 1.4E+00& &&&& C\\
\\
\multicolumn{8}{c}{$\mathrm{3p\,3d~^3D^o_3}	$ -- $\mathrm{3d^2~^3F_4}	$} \\
\textbf{VV~RCI}	                  & 264.09& 1.77897E+00& 1.86728E+00& 6.83E-01& 4.99E-01&   5.90906E+01 & B+  \\
\textbf{VV+CV+C~RCI int}          & 263.42& 1.71766E+00& 1.70682E+00& 6.36E-01& 2.63E-01&$-$4.46079E+02 & AA  \\
\textbf{VV+CV+C~RCI}              & 263.42& 1.71761E+00& 1.70689E+00& 6.36E-01& 2.63E-01&$-$4.51378E+02 & AA  \\
\textbf{VV+CV+C+CC~RCI int}       & 263.53& 1.72784E+00& 1.73836E+00& 6.41E-01& 2.68E-01&   4.66749E+02 & AA  \\
\textbf{VV+CV+C+CC~RCI}	          & 263.53& 1.72755E+00& 1.73736E+00& 6.41E-01& 2.68E-01&   5.00049E+02 & AA  \\
\textbf{RCI~(RSMBPT) 100\%}       & 263.53& 1.72767E+00& 1.73759E+00& 6.41E-01& 2.68E-01&   4.94699E+02 & AA  \\
\textbf{RCI~(RSMBPT) 100\% int}	  & 263.53& 1.72784E+00& 1.73834E+00& 6.41E-01& 2.68E-01&   4.67579E+02 & AA  \\
\textbf{RCI~(RSMBPT) 99.95\%}     & 263.54& 1.72770E+00& 1.73812E+00& 6.41E-01& 2.68E-01&   4.71248E+02 & AA  \\
\textbf{RCI~(RSMBPT) 99.95\% int} & 263.54& 1.72785E+00& 1.73874E+00& 6.41E-01& 2.68E-01&   4.50894E+02 & AA  \\
\textbf{RCI~(RSMBPT) 99.5\%}      & 263.58& 1.72787E+00& 1.73975E+00& 6.42E-01& 2.70E-01&   4.13263E+02 & AA  \\
\textbf{RCI~(RSMBPT) 99.5\% int}  & 263.58& 1.72798E+00& 1.74031E+00& 6.42E-01& 2.71E-01&   3.98493E+02 & AA  \\
\textbf{RCI~(RSMBPT) 99\%}	      & 263.58& 1.72811E+00& 1.73756E+00& 6.42E-01& 2.71E-01&   5.19328E+02 & AA  \\
\textbf{RCI~(RSMBPT) 99\% int}	  & 263.58& 1.72822E+00& 1.73814E+00& 6.42E-01& 2.72E-01&   4.94635E+02 & AA  \\
\textbf{RCI~(RSMBPT) 95\%}	      & 263.61& 1.72863E+00& 1.73934E+00& 6.43E-01& 2.75E-01&   4.58395E+02 & AA  \\
\textbf{RCI~(RSMBPT) 95\% int}	  & 263.60& 1.72870E+00& 1.73975E+00& 6.43E-01& 2.75E-01&   4.44366E+02 & AA  \\
NIST \cite{NIST_ASD}            &263.685 & & &&&& \\
\\
\multicolumn{8}{c}{$\mathrm{3p\,3d~^1F^o_3}	$ -- $\mathrm{3d^2~^1G_4}	$} \\
\textbf{VV~RCI}	                  &  291.65& 2.47785E+00& 2.50553E+00& 7.86E-01& 3.61E-01&   2.55295D+02 & A+  \\
\textbf{VV+CV+C~RCI int}          &  289.94& 2.39937E+00& 2.39099E+00& 7.42E-01& 2.20E-01&$-$8.08404D+02 & AA  \\
\textbf{VV+CV+C~RCI}              &  289.93& 2.39926E+00& 2.39116E+00& 7.42E-01& 2.20E-01&$-$8.35282D+02 & AA  \\
\textbf{VV+CV+C+CC~RCI int}       &  290.18& 2.41917E+00& 2.47806E+00& 7.51E-01& 2.28E-01&   1.18305D+02 & A  \\
\textbf{VV+CV+C+CC~RCI}	          &  290.18& 2.41869E+00& 2.47617E+00& 7.50E-01& 2.28E-01&   1.21141D+02 & A  \\
\textbf{RCI~(RSMBPT) 100\%}       &  290.18& 2.41894E+00& 2.47673E+00& 7.50E-01& 2.28E-01&   1.20497D+02 & A  \\
\textbf{RCI~(RSMBPT) 100\% int}	  &  290.18& 2.41918E+00& 2.47796E+00& 7.51E-01& 2.28E-01&   1.18529D+02 & A  \\
\textbf{RCI~(RSMBPT) 99.95\%}     &  290.19& 2.41910E+00& 2.47727E+00& 7.50E-01& 2.28E-01&   1.19762D+02 & A  \\
\textbf{RCI~(RSMBPT) 99.95\% int} &  290.19& 2.41924E+00& 2.47797E+00& 7.51E-01& 2.28E-01&   1.18630D+02 & A  \\
\textbf{RCI~(RSMBPT) 99.5\%}      &  290.22& 2.41940E+00& 2.47670E+00& 7.51E-01& 2.30E-01&   1.21529D+02 & A  \\
\textbf{RCI~(RSMBPT) 99.5\% int}  &  290.22& 2.41950E+00& 2.47729E+00& 7.51E-01& 2.30E-01&   1.20526D+02 & A  \\
\textbf{RCI~(RSMBPT) 99\%}	      &  290.23& 2.41970E+00& 2.47552E+00& 7.51E-01& 2.30E-01&   1.24713D+02 & A  \\
\textbf{RCI~(RSMBPT) 99\% int}	  &  290.23& 2.41979E+00& 2.47612E+00& 7.51E-01& 2.30E-01&   1.23607D+02 & A  \\
\textbf{RCI~(RSMBPT) 95\%}	      &  290.26& 2.42061E+00& 2.48296E+00& 7.51E-01& 2.32E-01&   1.11922D+02 & A  \\
\textbf{RCI~(RSMBPT) 95\% int}	  &  290.26& 2.42065E+00& 2.48329E+00& 7.51E-01& 2.32E-01&   1.11417D+02 & A  \\
NIST \cite{NIST_ASD}            &290.239 & & &&&& \\
\\
\multicolumn{8}{c}{$\mathrm{3s^2~^1S_0}	$ -- $\mathrm{3s\,3p~^1P^o_1}	$} \\
\textbf{VV~RCI}	                  & 281.82& 7.56430E-01& 7.73811E-01& 6.96E-01& 9.16E-01& 1.25218E+02  & A  \\
\textbf{VV+CV+C~RCI int}          & 283.99& 7.29184E-01& 7.38054E-01& 6.49E-01& 3.85E-01& 2.34642E+02  & A+  \\
\textbf{VV+CV+C~RCI}              & 283.99& 7.29185E-01& 7.38056E-01& 6.49E-01& 3.85E-01& 2.34601E+02  & A+  \\
\textbf{VV+CV+C+CC~RCI int}       & 284.17& 7.31169E-01& 7.40656E-01& 6.51E-01& 3.87E-01& 2.20107E+02  & A+  \\
\textbf{VV+CV+C+CC~RCI}	          & 284.17& 7.31146E-01& 7.40965E-01& 6.51E-01& 3.87E-01& 2.12736E+02  & A+  \\
\textbf{RCI~(RSMBPT) 100\%}       & 284.17& 7.31157E-01& 7.40734E-01& 6.51E-01& 3.87E-01& 2.18044E+02  & A+  \\
\textbf{RCI~(RSMBPT) 100\% int}	  & 284.17& 7.31169E-01& 7.40655E-01& 6.51E-01& 3.87E-01& 2.20131E+02  & A+  \\
\textbf{RCI~(RSMBPT) 99.95\%}     & 284.17& 7.31158E-01& 7.40735E-01& 6.51E-01& 3.87E-01& 2.18053E+02  & A+  \\
\textbf{RCI~(RSMBPT) 99.95\% int} & 284.17& 7.31170E-01& 7.40660E-01& 6.51E-01& 3.87E-01& 2.20060E+02  & A+  \\
\textbf{RCI~(RSMBPT) 99.5\%}      & 284.18& 7.31194E-01& 7.41184E-01& 6.51E-01& 3.88E-01& 2.09130E+02  & A+  \\
\textbf{RCI~(RSMBPT) 99.5\% int}  & 284.18& 7.31206E-01& 7.41113E-01& 6.51E-01& 3.88E-01& 2.10878E+02  & A+  \\
\textbf{RCI~(RSMBPT) 99\%}	      & 284.19& 7.31229E-01& 7.41656E-01& 6.51E-01& 3.89E-01& 2.00463E+02  & A+  \\
\textbf{RCI~(RSMBPT) 99\% int}	  & 284.19& 7.31241E-01& 7.41586E-01& 6.51E-01& 3.89E-01& 2.02046E+02  & A+  \\
\textbf{RCI~(RSMBPT) 95\%}	      & 284.19& 7.31539E-01& 7.41170E-01& 6.52E-01& 3.95E-01& 2.16958E+02  & A+  \\
\textbf{RCI~(RSMBPT) 95\% int}	  & 284.19& 7.31549E-01& 7.41128E-01& 6.52E-01& 3.95E-01& 2.18128E+02  & A+  \\
NIST \cite{NIST_ASD}            &284.164 & 7.75E-01& &&&& B \\
\\
\multicolumn{8}{c}{$\mathrm{3s^2~^1S_0}	$ -- $\mathrm{3s\,3p~^3P^o_1}	$} \\
\textbf{VV~RCI}	                  & 418.51& 4.45691E-03& 4.76312E-03& 4.90E-03& 3.14E-03& 4.32771E+01 & B+  \\
\textbf{VV+CV+C~RCI int}          & 417.21& 4.62691E-03& 4.88159E-03& 5.00E-03& 1.91E-03& 5.34981E+01 & B+  \\
\textbf{VV+CV+C~RCI}              & 417.21& 4.62689E-03& 4.88153E-03& 5.00E-03& 1.91E-03& 5.35056E+01 & B+  \\
\textbf{VV+CV+C+CC~RCI int}       & 417.76& 4.61939E-03& 4.88880E-03& 5.00E-03& 1.92E-03& 5.06082E+01 & B+  \\
\textbf{VV+CV+C+CC~RCI}	          & 417.75& 4.62086E-03& 4.89484E-03& 5.00E-03& 1.92E-03& 4.98157E+01 & B+  \\
\textbf{RCI~(RSMBPT) 100\%}       & 417.76& 4.61967E-03& 4.88114E-03& 5.00E-03& 1.92E-03& 5.20843E+01 & B+  \\
\textbf{RCI~(RSMBPT) 100\% int}	  & 417.76& 4.61935E-03& 4.88875E-03& 5.00E-03& 1.92E-03& 5.06101E+01 & B+  \\
\textbf{RCI~(RSMBPT) 99.95\%}     & 417.75& 4.61863E-03& 4.88024E-03& 5.00E-03& 1.92E-03& 5.20462E+01 & B+  \\
\textbf{RCI~(RSMBPT) 99.95\% int} & 417.75& 4.61834E-03& 4.88798E-03& 5.00E-03& 1.92E-03& 5.05559E+01 & B+  \\
\textbf{RCI~(RSMBPT) 99.5\%}      & 417.75& 4.60860E-03& 4.83438E-03& 4.99E-03& 1.91E-03& 5.98457E+01 & B+  \\
\textbf{RCI~(RSMBPT) 99.5\% int}  & 417.75& 4.60837E-03& 4.84216E-03& 4.99E-03& 1.91E-03& 5.78640E+01 & B+  \\
\textbf{RCI~(RSMBPT) 99\%}	      & 417.77& 4.61154E-03& 4.83255E-03& 4.99E-03& 1.91E-03& 6.11318E+01 & B+  \\
\textbf{RCI~(RSMBPT) 99\% int}	  & 417.77& 4.61139E-03& 4.84048E-03& 4.99E-03& 1.91E-03& 5.90466E+01 & B+  \\
\textbf{RCI~(RSMBPT) 95\%}	      & 417.71& 4.60610E-03& 4.77256E-03& 4.99E-03& 1.91E-03& 8.03799E+01 & B+  \\
\textbf{RCI~(RSMBPT) 95\% int}	  & 417.71& 4.60598E-03& 4.77933E-03& 4.99E-03& 1.91E-03& 7.72679E+01 & B+  \\
NIST \cite{NIST_ASD}            &417.258 & 4.4E-03& &&&& E \\
\\
\multicolumn{8}{c}{$\mathrm{3s\,3p~^3P^o_1}	$ -- $\mathrm{3p^2~^1D_2}	$} \\
\textbf{VV~RCI}	                  & 311.58& 8.23087E-02& 8.38464E-02& 4.22E-02& 3.80E-02& 1.53516E+02 & A+  \\
\textbf{VV+CV+C~RCI int}          & 312.30& 9.14394E-02& 9.21971E-02& 4.68E-02& 2.51E-02& 3.43440E+02 & AA  \\
\textbf{VV+CV+C~RCI}              & 312.30& 9.14393E-02& 9.21978E-02& 4.68E-02& 2.51E-02& 3.43121E+02 & AA  \\
\textbf{VV+CV+C+CC~RCI int}       & 311.95& 8.99187E-02& 9.02371E-02& 4.58E-02& 2.44E-02& 8.00907E+02 & AA  \\
\textbf{VV+CV+C+CC~RCI}	          & 311.95& 8.99279E-02& 9.02746E-02& 4.58E-02& 2.44E-02& 7.35835E+02 & AA  \\
\textbf{RCI~(RSMBPT) 100\%}       & 311.95& 8.99269E-02& 9.02691E-02& 4.58E-02& 2.44E-02& 7.45469E+02 & AA  \\
\textbf{RCI~(RSMBPT) 100\% int}	  & 311.95& 8.99186E-02& 9.02377E-02& 4.58E-02& 2.44E-02& 7.99079E+02 & AA  \\
\textbf{RCI~(RSMBPT) 99.95\%}     & 311.96& 8.99048E-02& 9.02339E-02& 4.58E-02& 2.44E-02& 7.74715E+02 & AA  \\
\textbf{RCI~(RSMBPT) 99.95\% int} & 311.96& 8.98970E-02& 9.02085E-02& 4.58E-02& 2.44E-02& 8.18511E+02 & AA  \\
\textbf{RCI~(RSMBPT) 99.5\%}      & 311.98& 8.98396E-02& 9.03289E-02& 4.58E-02& 2.44E-02& 5.21358E+02 & AA  \\
\textbf{RCI~(RSMBPT) 99.5\% int}  & 311.98& 8.98320E-02& 9.02995E-02& 4.58E-02& 2.44E-02& 5.45616E+02 & AA  \\
\textbf{RCI~(RSMBPT) 99\%}	      & 311.97& 8.97719E-02& 9.04057E-02& 4.57E-02& 2.45E-02& 4.02754E+02 & AA  \\
\textbf{RCI~(RSMBPT) 99\% int}	  & 311.96& 8.97646E-02& 9.03791E-02& 4.57E-02& 2.45E-02& 4.15255E+02 & AA  \\
\textbf{RCI~(RSMBPT) 95\%}	      & 311.99& 8.97523E-02& 9.00675E-02& 4.58E-02& 2.48E-02& 8.07672E+02 & AA  \\
\textbf{RCI~(RSMBPT) 95\% int}	  & 311.99& 8.97460E-02& 9.00471E-02& 4.58E-02& 2.48E-02& 8.45346E+02 & AA  \\
NIST \cite{NIST_ASD}            &312.556 & 8.3E-02& &&&& E \\
\noalign{\smallskip}
\end{longtable}
}

\section{Conclusions}
The method based on the Rayleigh-Schr\"odinger perturbation theory in an irreducible tensorial form
is extended to take into account the CC correlations, providing the expressions for the estimation of these correlations.
This extended RSMBPT method allows to estimate the contribution 
of any $K'$ configuration of the CV, C and CC correlation 
with preferred core and virtual orbitals sets for any atom or ion. 

The combination of the RCI and the RSMBPT methods works perfectly 
for energy and transition parameters calculations, when the CV, C and CC correlations 
are included using the RSMBPT method, as it is shown in the examples.
The developed method has an advantage over the regular method because it allows the selection of 
the most relevant CV, C, CC correlations and significantly reduces the CSF space. 
Furthermore, the RSMBPT method allows the inclusion of correlations that cannot be included 
using the regular {\sc Grasp}2018 calculation scheme.
The above mentioned advantages of the RSMBPT method over the regular method would be 
useful and beneficial for calculations involving complex atoms and ions.

The change in CFs was observed when a new group of correlations was added to the regular {\sc Grasp}2018  calculations.
Meanwhile using the RSMBPT method, the values of CF are stable when the most important configurations 
of CV, C, and CC correlations with the different amount of these correlations are included. This 
indicates that the most important correlations are included in the calculations and the line strengths would not change.
Thus, the QQE method should provide a more accurate estimate of the uncertainties when the RSMBPT method is used.
This is another advantage of the developed method over the regular {\sc Grasp}2018 approach.

\bibliographystyle{pccp}

\end{document}